\documentclass[review]{elsarticle}
\UseRawInputEncoding
\usepackage{lineno, hyperref}
\modulolinenumbers[5]

\journal{Journal of \LaTeX\ Templates}

\usepackage{array, makecell} 
\usepackage{lscape}
\usepackage{multirow} 
\usepackage{graphicx} 
\usepackage{rotating}
\usepackage{longtable, booktabs, tabularx, enumitem, ragged2e} 
\newcolumntype{L}{>{\raggedright\arraybackslash}X}
\usepackage[margin=25mm]{geometry}
\usepackage{ltablex}
\usepackage{amsmath}
\usepackage[tableposition=t]{caption} 
\usepackage{comment}
\usepackage{soul}
\newcolumntype{C}[1]{>{\centering\arraybackslash}m{#1}}
\usepackage{amssymb} 
\usepackage[table]{xcolor}
\definecolor{mediumblue}{rgb}{0.0, 0.0, 0.8}

\usepackage{tcolorbox}

\begin{document}
\begin{frontmatter}

\title{\textbf{Software Security Patch Management - A Systematic Literature Review of Challenges, Approaches, Tools and Practices}}

\author[mymainaddress,mysecondaryaddress]{Nesara Dissanayake\corref{correspondingauthor}}
\ead{nesara.madugodasdissanayakege@adelaide.edu.au}
\author[mymainaddress,mysecondaryaddress]{Asangi Jayatilaka} \ead{asangi.jayatilaka@adelaide.edu.au}
\author[mymainaddress,mysecondaryaddress]{Mansooreh Zahedi} \ead{mansooreh.zahedi@adelaide.edu.au}
\author[mymainaddress,mysecondaryaddress]{M. Ali Babar} \ead{ali.babar@adelaide.edu.au}

\cortext[correspondingauthor]{Corresponding author.}

\address[mymainaddress]{School of Computer Science, The University of Adelaide, Australia}
\address[mysecondaryaddress]{Centre for Research on Engineering Software Technology (CREST), Australia}

\begin{abstract}

\emph{\textbf{Context:}} Software security patch management purports to support the process of patching known software security vulnerabilities. Patching security vulnerabilities in large and complex systems is a hugely challenging process that involves multiple stakeholders making several interdependent technological and socio-technical decisions. Given the increasing recognition of the importance of software security patch management, it is important and timely to systematically review and synthesise the relevant literature on this topic.

\emph{\textbf{Objective:}} This paper aims at systematically reviewing the state of the art of software security patch management to identify the socio-technical challenges in this regard, reported solutions (i.e., approaches, tools, and practices), the rigour of the evaluation and the industrial relevance of the reported solutions, and to identify the gaps for future research.

\emph{\textbf{Method:}} We conducted a systematic literature review of 72 studies published from 2002 to March 2020, with extended coverage until September 2020 through forward snowballing. 

\emph{\textbf{Results:}} We identify 14 socio-technical challenges in software security patch management, 18 solution approaches, tools and practices mapped onto the software security patch management process. We provide a mapping between the solutions and challenges to enable a reader to obtain a holistic overview of the gap areas. The findings also reveal that only 20.8\% of the reported solutions have been rigorously evaluated in industrial settings.

\emph{\textbf{Conclusion:}} Our results reveal that 50\% of the common challenges have not been directly addressed in the solutions and that most of them (38.9\%) address the challenges in one phase of the process, namely vulnerability scanning, assessment and prioritisation. Based on the results that highlight the important concerns in software security patch management and the lack of solutions, we recommend a list of future research directions. This study also provides useful insights about different opportunities for practitioners to adopt new solutions and understand the variations of their practical utility. 

\end{abstract}

\begin{keyword}
    Security patch management \sep%
    Vulnerability management\sep%
    Systematic literature review
\end{keyword}
\end{frontmatter}

\section{Introduction}
\label{section:introduction}
Cyber attackers exploiting software vulnerabilities remain one of the most critical risks facing modern corporate networks. Patching continues to be the most effective and widely recognised strategy for protecting software systems against such cyberattacks. However, despite the rapid releases of security patches addressing newly discovered vulnerabilities in software products, a majority of cyberattacks in the wild have been a result of an exploitation of a known vulnerability for which a patch had already existed \cite{AccentureReport2020, UnpatchedChallenges, CyberSecurityReview, EquifaxBlame}. More disturbing is that such cyberattacks have caused devastating consequences resulting in huge financial and reputation losses from breach of confidentiality and integrity of company data such as the Equifax case \cite{ForbesEquifaxLoss, EquifaxNoExcuse}, and even human death due to the unavailability of software systems \cite{GermanCyberAttack2020}. These outcomes can be largely attributed to the inherent complex issues facing when applying software security patches in organisational IT environments.

Software security patch management refers to the process of applying patches to the security vulnerabilities present in the software products and systems deployed in an organisation’s IT environment. The process consists of identifying existing vulnerabilities in managed software systems, acquiring, testing, installing, and verifying software security patches \cite{souppaya2013guide, li2019keepers, tiefenau2020security, gentile2019survey}. Performing these activities involve managing interdependencies between multiple stakeholders, and several technical and socio-technical tasks and decisions that make software security patch management a complex issue. This issue is exacerbated by the need to balance between applying a software security patch as early as possible and having to patch a myriad of enterprise applications, often leaving a large number of unpatched systems vulnerable to attacks. According to a recent industry report \cite{Automox}, more than 50\% of organisations are unable to patch critical vulnerabilities within the recommended time of 72 hours of their release, and around 15\% of them remain unpatched even after 30 days. Such evidence reveals that modern organisations are struggling to meet the requirements of ``patch early and often" indicating an increasing need for more attention to the software security patching process in practice. 

Despite the criticality of software security patch management in the industry, this is still an emerging area of rising interest in research that needs further attention. While there have been many studies aimed at providing technical advancements to improve software security patch management tasks (e.g., an algorithm for optimising patching policy selection \cite{dey2015optimal}), the socio-technical aspects of software security patch management have received relatively limited attention \cite{dissanayake2021grounded}. Socio-technical aspects relate to the organisational process, policies, skill and resource management, and interaction of people with technical solutions \cite{islam2019multi}. This is an important limitation because the software security patch management process is inherently a socio-technical endeavour, where human and technological interactions are tightly coupled, such that the success of software security patch management significantly depends on the effective collaboration of humans with the technical systems. The understanding of the socio-technical aspects is essential for identifying the prevailing issues and improving the effectiveness of the software security patch management process \cite{dissanayake2021grounded, li2019keepers, tiefenau2020security}. To the best of our knowledge, there has been no review or survey aimed at organising the body of knowledge on the socio-technical aspects of software security patch management covering the existing challenges and solutions in this area.

Responding to this evident lack of attention to a highly critical and timely topic with growing interest in the industry and academia, we aimed at systematically analysing the literature on software security patch management. A systematic review would largely benefit both researchers and practitioners to gain an in-depth holistic overview of the state-of-the-art of software security patch management as well support in transferring the research outcomes to industrial practice \cite{kitchenham2004evidence}. Further, the results provide useful insights to identify the limitations of the existing solutions, and gaps that need the attention of the research community. We focus on the less-explored socio-technical aspects of software security patch management investigating the challenges and existing solutions to address those challenges reported in 72 primary studies. The key contributions of this novel systematic literature review (SLR) are as follows:

\begin{itemize}
    \item A consolidated body of knowledge of research on socio-technical aspects of software security patch management providing guidance for practitioners and researchers who want to better understand the process.
    \item A comprehensive understanding of the socio-technical challenges faced in the security patching process.
    \item A classification of the current solutions in terms of approaches, tools, and practices to address the challenges and mapping of the challenges to the proposed solutions.
    \item An analysis of the proposed solutions' rigour of evaluation and industrial relevance to inform and support the transferability of research outcomes to an industrial environment.
    \item Identification of the potential gaps for future research highlighting important and practical concerns in software security patch management that require further attention.
\end{itemize}

The rest of this paper is organised as follows. Section \ref{section:background} describes the background and other reviews related to the topic. Section \ref{section:method} describes the research methodology used for this SLR. Sections \ref{section:challenges}, \ref{section:approachesandtools}, \ref{section:practices} and \ref{section:evaluation} present the findings to the research questions. In Section \ref{section:discussion}, we discuss key future research, and threats to the validity are presented in Section \ref{section:threats}. Finally, Section \ref{section:conclusion} concludes the review.

\section{Background and Related Work}
\label{section:background}

In this section, we present an overview of the software security patch management process. Then, we provide a comparison of our study with the existing related reviews.

\subsection{Overview of the software security patch management process}

Given there is no commonly known/accepted definition of software security patch management, we decided to devise an operational definition for our research based on our evidence-based understanding from a longitudinal case study \cite{dissanayake2021grounded} and the existing related literature \cite{souppaya2013guide, mell2005creating, gerace2005critical, brykczynski2003reducing, nicastro2003security, dey2015optimal, cavusoglu2008security, cavusoglu2006economics}. Hence, our operational definition of software security patch management is below:

\begin{tcolorbox}
    \textbf{Software security patch management is a multifaceted process of identifying, acquiring, testing, installing, and verifying security patches for software products and systems.}
\end{tcolorbox}

\begin{figure}[h!]
    \centering
    \begin{minipage}{0.9\textwidth}
    \centering
    \includegraphics[width=0.9\linewidth]{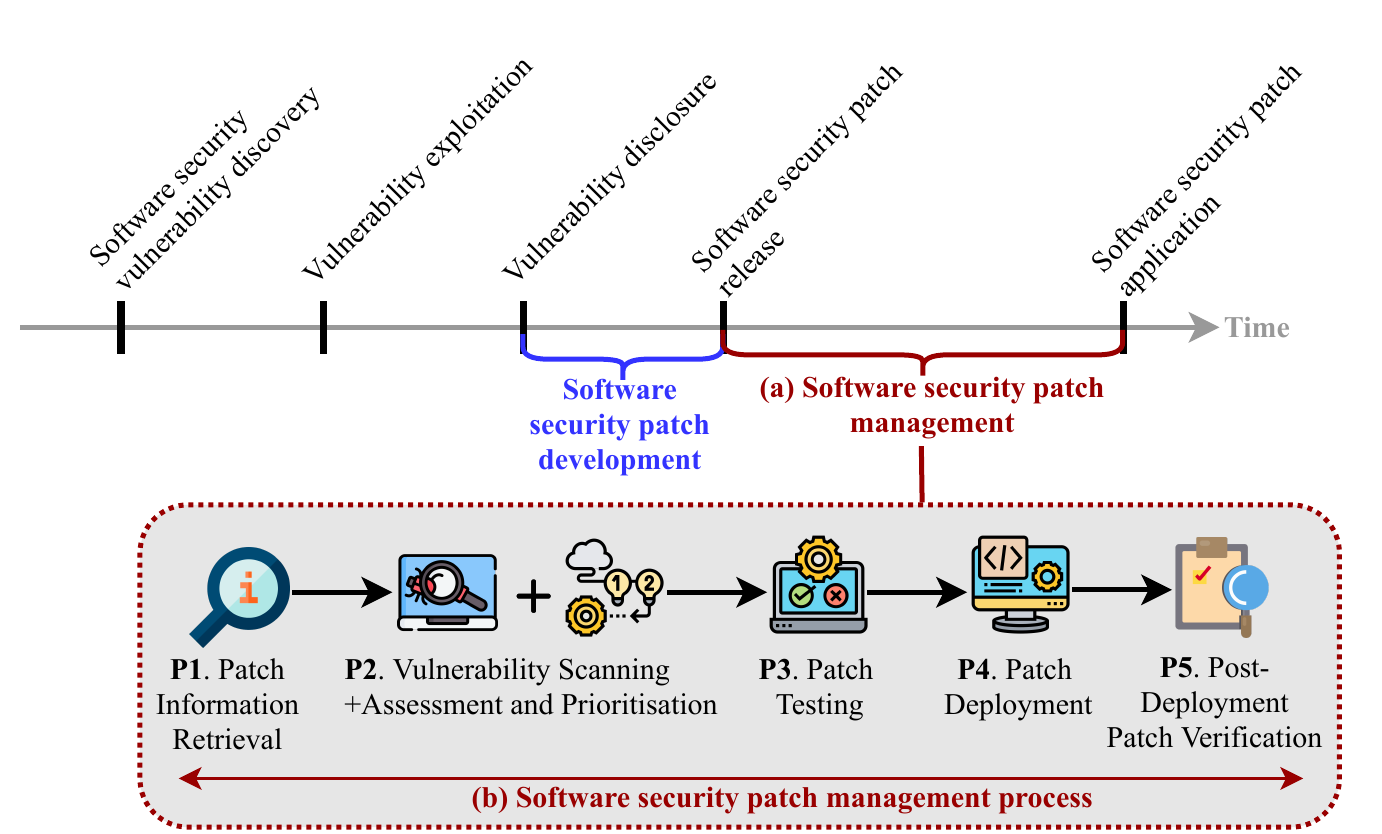}
    \captionof{figure}{(a)~The focus of software security patch management in the software vulnerability life cycle. (b)~An overview of the software security patch management process.}
    \label{fig:slrfocus}
    \end{minipage}
\end{figure}

Software security patch management is a security practice designed to proactively prevent the exploitation of security vulnerabilities that exist within an organisation's deployed software products and systems \cite{nicastro2003security, mell2005creating}. Software security patches are \textit{``pieces of code developed to address security problems identified in software"} \cite{mell2005creating}. In general, software security patches are always prioritised over non-security patches by industry practitioners and researchers as they are aimed at mitigating software vulnerabilities (or security bugs) that present exploitable opportunities for malicious entities to gain access to systems \cite{mell2005creating, souppaya2013guide}. In addition, software security patches are acknowledged as the most effective strategy to mitigate software vulnerabilities \cite{souppaya2013guide, araujo2014patches}. A successful software security patch management process is thus essential and critical to sustaining the confidentiality, integrity, and availability of IT systems \cite{mell2005creating}. Figure \ref{fig:slrfocus}(a) shows the focus of software security patch management from a typical software vulnerability life cycle perspective. For example, the focus here relates to the process of a company (company A) applying security patches to its deployed third-party software after the patches are released by the corresponding third-party software vendors (company B).

Despite the importance, software security patch management remains one of the most challenging efforts facing IT practitioners. Figure \ref{fig:slrfocus}(b) illustrates the five main phases of the software security patch management process \cite{souppaya2013guide, li2019keepers, tiefenau2020security}. Firstly, in the \textit{patch information retrieval} phase, practitioners learn about new patches and acquire them from third-party software vendors like Microsoft. In the next phase of \textit{vulnerability scanning, assessment and prioritisation}, the practitioners scan the managed software systems for the newly disclosed vulnerabilities to identify the applicability of patches in their organisational context, assess the risk, and correspondingly prioritise the patching decisions. Followed by the \textit{patch testing} phase whereby the patches are tested for accuracy and stability and prepared for installation by changing machine configurations, resolving patch dependencies and making backups. Then the patches are installed at their target machines in the \textit{patch deployment} phase. Finally, the patch deployments are verified through monitoring (for unexpected service interruptions) and post-deployment issues are handled in the \textit{post-deployment patch verification} phase. 


\subsection{Other reviews related to software security patch management}

Despite the increasing demand and growing body of literature on the topic of software security patch management, we did not find any existing systematic literature review or systematic mapping study focused on the software security patch management process. However, there have been several existing reviews/surveys on software security patch management-related topics (Table \ref{tab:relatedwork}), for example, international standards on patch management and dynamic software updating (DSU) \cite{hicks2005dynamic}, i.e., a method that allows for runtime patching without restarts or downtime. A comparison between these existing reviews/surveys on related topics and our SLR is presented below.

Recently, Gentile and Serio \cite{gentile2019survey} reviewed a set of existing international standards on patch management and current industry best practices, assessing their relevance to the context of complex and critical infrastructures, particularly the industrial control systems (ICSs). Based on the survey results, they defined a general-purpose workflow to support the patch management process in the ICSs. While our study provides a set of practices for successful security patch management similar to this survey study, we also include a set of recommendations, guidelines, lessons learned and shared experiences of researchers and industry practitioners providing more coverage. However, the main difference between the study conducted by Gentile and Serio \cite{gentile2019survey} and our study is the research focus, i.e., our SLR focuses on the existing challenges and solutions in software security patch management which is not covered in the study by Gentile and Serio \cite{gentile2019survey} that is limited to ICS context.

The other set of review and mapping studies exclusively focus on dynamic software updating (DSU) \cite{hicks2005dynamic}. DSU aims at live patching to avoid restarts or downtime that cause service interruptions. As such, the contributions of these studies focus on facilitating one phase in the patch management process, namely the patch deployment phase (Figure \ref{fig:slrfocus}(b)). For example, Miedes et al. \cite{miedes2012dynamic} in their technical report surveyed and classified the common dynamic update mechanisms providing an overview of the concepts and techniques of DSU in the literature. Subsequently, several studies \cite{seifzadeh2013survey, gregersen2013state, mugarza2018analysis, ahmed2020dynamic} followed the trend reviewing the state-of-the-art of DSU techniques such as the review by Seifzadeh et al. \cite{seifzadeh2013survey} in which they provided a framework for evaluating the DSU features. Moreover, they highlight the need for future research investigating the challenges of adopting runtime patching in organisations that have been investigated in our study. In summary, our review differs from the existing studies by contributing to the gap area of a lack of a systematic review that identifies and analyses the challenges and solutions in software security patch management.

\begin{table}[h!]
\centering
\caption{Comparison of contributions between our study and the existing related reviews/surveys.}
\label{tab:relatedwork}
\footnotesize\sffamily
  \begin{tabular}{|>{\raggedright\arraybackslash}m{.12\textwidth} | >{\raggedright\arraybackslash}m{.28\textwidth} | >{\raggedright\arraybackslash}m{.2\textwidth} | C{.07\textwidth} | C{.1\textwidth} | C{.08\textwidth} |}
    \hline
    \centering \textbf{Study} & \centering \textbf{Study contribution} & \centering \textbf{Focus on software security patch management} & \centering \textbf{Challenges} & \textbf{Solutions} & \textbf{Evaluation of solutions}  \\
    \hline
    Gentile and Serio 2019 \cite{gentile2019survey} & International standards and best practices for patch management of complex industrial control systems & Overall software patch management process & -- & \checkmark (ICS specific best practices) & --  \\
    \hline
    Miedes and Munoz-Escoi 2012 \cite{miedes2012dynamic} & A classification of the dynamic software update (DSU) mechanisms & Patch deployment phase & -- & \checkmark (DSU types) & -- \\
    \hline
    Seifzadeh et al. 2013 \cite{seifzadeh2013survey} & A framework for the evaluation of dynamic updating features & Patch deployment phase & -- & -- & \checkmark (DSU features) \\
    \hline
    Gregersen et al. 2013 \cite{gregersen2013state} & A systematic mapping of DSU approaches, tools, models, and techniques & Patch deployment phase & -- & -- & \checkmark (DSU features) \\
    \hline
    Mugarza et al. 2018 \cite{mugarza2018analysis} & An analysis of existing DSU techniques for industrial control systems & Patch deployment phase & -- & \checkmark (safety-compliant) & -- \\
    \hline
    Ahmed et al. 2020 \cite{ahmed2020dynamic} & A framework for the evaluation of dynamic updating features & Patch deployment phase & -- & \checkmark (DSU approaches, tools) & -- \\
    \hline
    \rowcolor{lightgray}
    This study & A SLR on the socio-technical aspects of software security patch management & Overall software security patch management process & \checkmark & \checkmark & \checkmark \\
    \hline
  \end{tabular}
\end{table}
 
\section{Research Methodology}
\label{section:method}

\begin{figure}[h!]
    \centering
    \begin{minipage}{0.95\textwidth}
    \centering
    \includegraphics[width=0.95\linewidth]{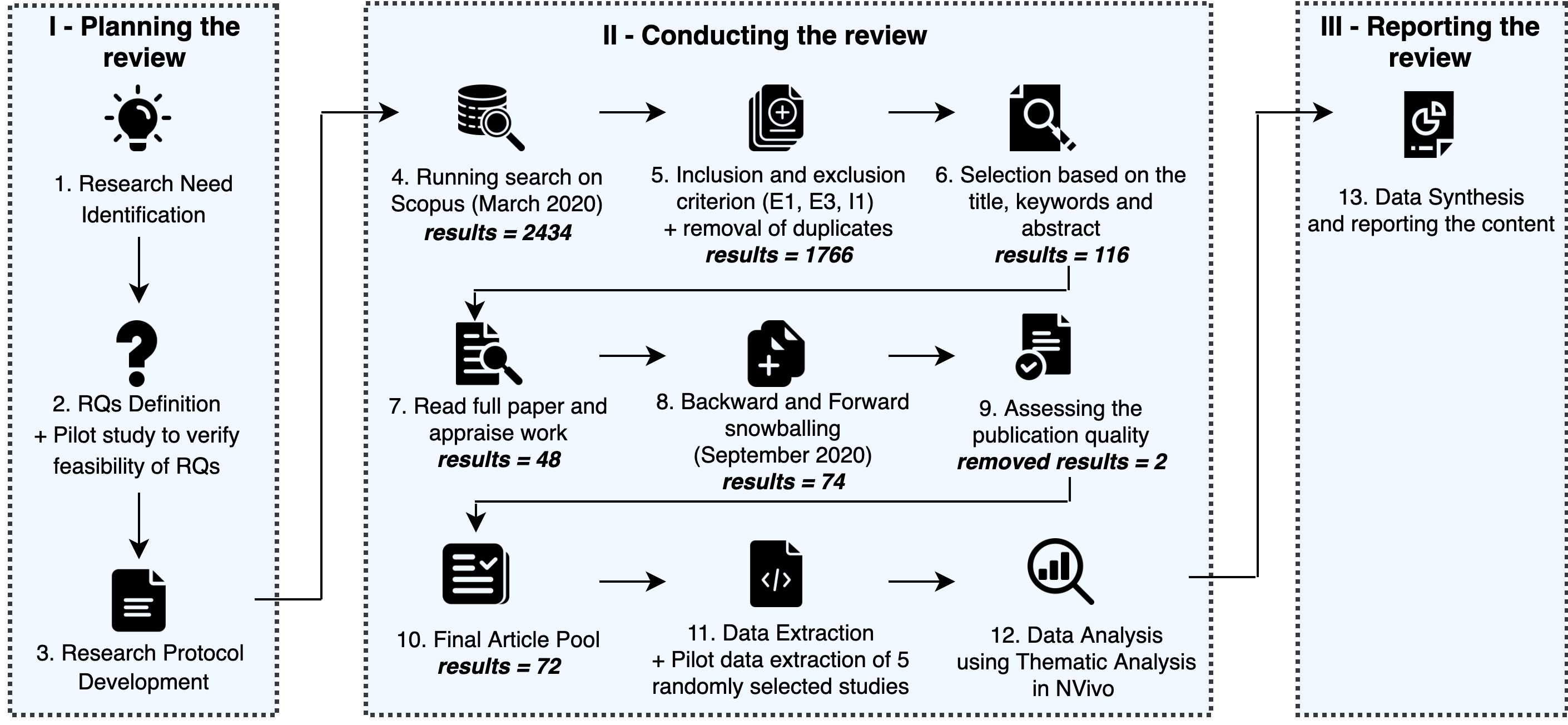}
    \captionof{figure}{An overview of the research methodology.}
    \label{fig:researchmethodology}
    \end{minipage}
\end{figure}

We conducted this SLR by following the guidelines proposed by Kitchenham and Charters  \cite{keele2007guidelines}. This section provides the details on the process illustrated in Figure \ref{fig:researchmethodology}. 

To summarise the execution process of this SLR, after all the authors discussed and agreed on the research questions, the first author developed a review protocol. All authors were involved in the search string construction following several pilot searches and multiple rounds of discussions. The study selection was jointly done by the first two authors and verified by the last author. The first author conducted a pilot data extraction (DE) where the DE form and quality assessment (QA) form were reviewed by all authors. Then, the first author performed data extraction, quality assessment and data synthesis under the close supervision of other authors who are experienced in conducting SLRs in SE. The data extraction and synthesis of results were regularly discussed and verified by all in weekly meetings throughout the process of 8 months.

\subsection{Research Questions}

This SLR is aimed at providing an overview of the ``state of the art'' in software security patch management. We formulated three research questions (RQs) to guide this SLR. Table \ref{tab:rq} presents the RQs, along with their motivations. 

\begin{table}[h!]
\centering
\caption{Research questions and their motivations}
\label{tab:rq}
\footnotesize\sffamily
\begin{tabular}{p{0.37\textwidth}p{0.63\textwidth}}
\toprule
\textbf{Research Question (RQ)}&\textbf{Motivation}\\
\midrule
\textbf{RQ1:} What socio-technical challenges have been reported in software security patch management? & This RQ aims to understand the socio-technical challenges faced by practitioners in the software security patch management process. \\
\addlinespace
\textbf{RQ2:} What types of solutions have been \newline{} proposed? \newline{} {RQ2.1. What approaches and tools have been proposed to facilitate software security patch management?} \newline{} {RQ2.2. What practices have been reported to successfully implement the software security patch management process?} & The motive of this question is to obtain a detailed understanding of the reported solutions in terms of approaches and tools (RQ2.1); and practices including industry experts' recommendations, guidelines, best practices, lessons learned and shared experiences for a successful software security patch management process (RQ2.2).\\
\addlinespace
\textbf{RQ3:} How have the solutions been assessed? \newline{} {RQ3.1. What types of evaluation have been used to assess the proposed solutions?} \newline{} {RQ3.2. What is the level of rigour and industrial relevance of the reported solutions?} & This RQ is aimed at analysing how the proposed solutions have been assessed. Since software security patch management is highly industry-centric, identifying the types of evaluation used to assess the proposed solutions (RQ3.1); and understanding how well the solutions have been evaluated aligned with industrial relevance (RQ3.2) would help practitioners to adopt the solutions, and researchers to understand the gaps in the current evaluation approaches. \\
\bottomrule
\end{tabular}
\end{table}

The answers to these RQs will provide an in-depth understanding of the socio-technical challenges in software security patch management (RQ1), available solutions (RQ2) and how the solutions have been evaluated (RQ3). The findings will enable researchers to identify gaps in this domain and potential future directions. It should be noted that we present solutions (RQ2) in two categories namely \textit{approaches and tools}, and \textit{practices}. We followed a similar strategy to Shahin et al. \cite{shahin2017continuous} in distinguishing between approaches and tools, and practices. The Cambridge dictionary defines an approach, method, and technique as \textit{``a particular way of doing something or handling a problem''}; a tool as \textit{``something that helps in a particular activity''}; and practice as \textit{``the act of doing something regularly or repeatedly''} \cite{Cambridgedictionary}. In this review, we define approach, along with framework, method, technique, and tool as \textit{a technical approach for addressing problems in software security patch management}, and classify them in the category of \textbf{\emph{``approaches and tools''}}, for ease of reference. It should be noted that we categorised the studies that provided a comparison overview of the existing tools also under \textit{``tools"}. Some studies reported more than one type of solution, hence those studies were included in more than one category. Practices, on the other hand, are defined as social practice and shared standards that can be supported by an approach or tool to facilitate a process \cite{dittrich2016does, schmidt2014concept}. We classified the recommendations, guidelines, best practices, lessons learned and shared experiences as \textbf{\emph{``practices''}} in this review study.

\subsection{Search Strategy}

We decided to use only the Scopus search engine to identify the relevant primary studies. The decision was based on the experiences reported by several other studies \cite{shahin2017continuous, kitchenham2010systematic, zahedi2016systematic, shahin2020architectural} justifying that Scopus indexes a large majority of the journals and conference papers in software engineering indexed by many other search engines, including ACM Digital Library, IEEE Xplore, Science Direct, Wiley Online Library and SpringerLink. Furthermore, there are several restrictions placed by the other digital libraries (e.g. SpringerLink, Wiley Online Library, IEEE Xplore) on large-scale searches on the meta-data of the published studies. Additionally, the search string needs to be modified for each single digital library that can result in errors being introduced. Therefore, running the search string on Scopus enabled us to use one search string while retrieving mostly relevant hits. We performed the search using the studies' title, abstract, and keywords. 

Initially, the search string was developed by selecting keywords based on the related literature and the reference lists from those relevant primary studies.
Then, we systematically modified it by adding a set of alternative search terms obtained through synonyms and subject headings used in the existing related research papers. The identified search terms were merged using Boolean AND and OR to construct various combinations of search strings. Based on these search string combinations, we conducted several pilot searches to find out the best search string and verify the inclusion of well-known primary studies. However, given that software security patch management is still a new and emerging topic in research, we observed inconsistent or use of different terminology in the literature. For example, one of the initial search terms was ``patch management" but we decided to exclude it as a keyword since the inclusion of the term ``management" resulted in returning a large number of irrelevant studies due to the inconsistent use of the term in the literature. A similar decision was followed for the keywords: \textit{``update"} and \textit{``socio-technical"}. Although these keywords were not included, the structure of the search string was capable of finding relevant papers, but we had to identify these papers through the study inclusion/exclusion phases. Furthermore, to mitigate the risk of missing potential primary studies from these decisions, we kept the search string as generic as possible and did not limit the search to any particular time range. Following the above strategies, we finalised the search string presented below: 
\\*

\noindent\fbox{%
    \parbox{\textwidth}{%
        \textit{\textbf{TITLE-ABS-KEY} ((`software' OR `system') AND (`patch*') AND (`security' OR `vulnerabilit*'))}
    }%
}
\\*


\subsection{Study Selection}

We retrieved 2434 studies from the execution of the search string on Scopus on 31st of March 2020. We did not restrict our search based on publication year. We applied snowballing \cite{wohlin2014guidelines} to scan the references of the selected studies to find more potential studies. A backward and forward search ensured the extensiveness of our snowballing results and extended the coverage of included studies until September 2020. 

We filtered the retrieved studies based on the inclusion and exclusion criteria presented in Table \ref{tab:iec}, initially defined when the review protocol was developed. Specifically, we included the studies related to the process of application of software security patches, that were in line with our SLR objectives and RQs. We refined the criteria during several iterations of search and study selection to ensure we achieve accurate classification of papers. For example, we did not include the short papers (E2) because they presented only concepts or ideas instead of well-defined, concrete approaches and did not provide sufficient and relevant evidence to answer the RQs. A similar strategy has been followed by several other SLR studies, for example, \cite{shahin2017continuous}. Similarly, we decided to introduce E3 during the pilot searches as we observed that those studies did not provide sufficient or useful data to extract based on the data extraction form. This decision was mutually agreed upon by all authors after a careful review of the full text in several rounds of the study selection process. Correspondingly, the application of the inclusion and exclusion criteria and assessment of publication quality (Section 3.4) resulted in a final count of 72 primary studies to be included in the data extraction, as listed in \ref{appendix:studieslist}. 

\begin{table}[h]
  \caption{Inclusion and exclusion criteria.}
  \label{tab:iec}
  \begin{footnotesize}
      \begin{tabular}{p{1\textwidth}}
        \toprule
        \textbf{Inclusion Criteria}\\
        \midrule
        \textbf{I1} Full text of peer-reviewed conference or journal article in English that is accessible.\\
        \textbf{I2} A study that relates to or addresses at least one phase of the software security patch management process (i.e. the phases in Figure \ref{fig:slrfocus}(b)).\\
        \hline
        \textbf{Exclusion Criteria}\\
        \hline
        \textbf{E1} Workshop articles, books, and non-peer-reviewed papers such as editorials, position papers, keynotes, reviews, tutorials, and panel discussions.\\
        \textbf{E2} Short papers (i.e., less than 6 pages).\\
        \textbf{E3} A study that reports only numerical analysis, algorithms, mathematical techniques related to software security patch management\\
        \textbf{E4} A study that is only focused on hardware or firmware.\\
        \textbf{E5} A study that is not in the domain of software security patch management (i.e. outside the focus area in Figure \ref{fig:slrfocus}(a)).\\
        \textbf{E6} Full text is unavailable.\\
        \bottomrule
        \end{tabular}
\end{footnotesize}
\end{table} 

\subsection{Assessing the publication quality}

We assessed the quality of the reviewed primary studies with regard to their ability to help answer the RQs and the effect on the drawn conclusions \cite{keele2007guidelines}. We developed a quality assessment criteria adopted and modified from a few published studies \cite{dybaa2008empirical, shahin2014systematic, garousi2019guidelines}. Table \ref{tab:aqp} provides the summary of the quality assessment. We graded the reviewed studies on each element of the quality assessment criteria using a three points (``Yes'', ``Partially'' or ``No'') scale. We assigned the values: 2 to Yes, 1 to Partially, and 0 to No, to produce a quantifiable result. A paper was considered to be of acceptable quality and therefore included in the SLR, if it received an average score $\geq$ 0.5. Two studies were excluded based on the quality assessment score. The first author performed the quality assessment while the second author validated the results by independently performing the quality assessment of a smaller set of randomly selected studies. Any disagreements were sorted through discussions. The quality assessment was used to exclude studies with low quality, and to indicate the credibility of the study’s findings \cite{dybaa2008empirical, Kitchenham07guidelinesfor}.


\begin{table}[h]
\centering
\caption{Assessment of the quality of publications.}
\label{tab:aqp}
\footnotesize\sffamily
\resizebox{\textwidth}{!}{%
    \begin{tabular}{p{0.05\textwidth}p{0.5\textwidth}p{0.15\textwidth}p{0.15\textwidth}p{0.15\textwidth}}
    \toprule
    \textbf{Id} & \textbf{Quality assessment criteria} & \textbf{Yes} & \textbf{Partially} & \textbf{No}\\
    \midrule
    C1 & Does the paper have clearly stated aims and objectives? & 63(87.5\%) & 9(12.5\%) & 0(0.0\%)\\
    C2 & Does the paper provide a clear context (e.g., industry or laboratory setting)? & 54(75\%) & 13(18.1\%) & 5(6.9\%)\\
    C3 & Does the paper have a research design that supports the aims? & 51(70.8\%) & 21(29.2\%) & 0(0.0\%)\\
    C4 & Does the paper explicitly discuss the limitations? & 22(30.6\%) & 9(12.5\%) & 41(56.9\%)\\
    C5 & Does the paper add value for research or practice of software security patch management? & 42(58.3\%) & 28(38.9\%) & 2(2.8\%)\\
    C6 & Does the paper provide a clear statement of findings? & 52(72.2\%) & 19(26.4\%) & 1(1.4\%)\\
    \bottomrule
    \end{tabular}}
\end{table}%

\subsection{Data Extraction}

We extracted data from the selected primary studies using a pre-defined data extraction (DE) form in an Excel spreadsheet as presented in \ref{appendix:dataextractionform}. The first author conducted a pilot DE on five randomly selected studies under the supervision of the other authors, and refined the DE form to capture all the required information in the best possible summarised version, through continuous discussions. We extracted some demographic information (e.g., authors name, venue published, and published year), and wrote critical summaries of the extracted data to be analysed and synthesised for answering the RQs.

\subsection{Data Analysis and Synthesis}

The demographic and contextual set of data items (D1 to D10 in \ref{appendix:dataextractionform}) were analysed using descriptive statistics while the other set of data items (D11 to D16 in \ref{appendix:dataextractionform}) were analysed using thematic analysis \cite{braun2006using, cruzes2011research}, a widely used qualitative data analysis method. The decision to use thematic analysis was based on our effort to classify the reported socio-technical challenges and solutions in the domain. We used the following steps guided by Braun and Clarke's thematic analysis process \cite{braun2006using} to synthesise the qualitatively gathered evidence. 

\textit{Familiarising with data}: First, we got familiarised with the extracted data by carefully reading each set of data. All data in the DE sheet were saved in the NVivo data analysis tool and shared among all authors.

\textit{Open coding}: Open coding started with breaking the data into smaller components to generate the initial codes. A code (i.e., a phrase) of three-five words was assigned to summarise the key points of the data.

\textit{Building themes}: Next, we assigned the codes to potential themes by iteratively revising and merging the codes based on their similarities using a multi-layered coding structure in NVivo. 

\textit{Merging themes}: Iteratively applying constant comparison on the codes and themes that emerged within one paper and between different papers, we grouped them to produce higher levels of themes. In the final step, we mapped the aggregated data to the software security patch management process phases based on the literature \cite{li2019keepers, tiefenau2020security}. The synthesis results for each RQ were carefully reviewed by all authors in weekly meetings before finalising the answers to the RQs.

\subsection{Overview of Selected Primary Studies}

In this subsection, we report the findings of the descriptive analysis of the demographic and contextual set of data items extracted.

\subsubsection{Demographic data}

\begin{figure}
    \centering
    \begin{minipage}{0.49\textwidth}
        \centering
        \includegraphics[width=1\textwidth]{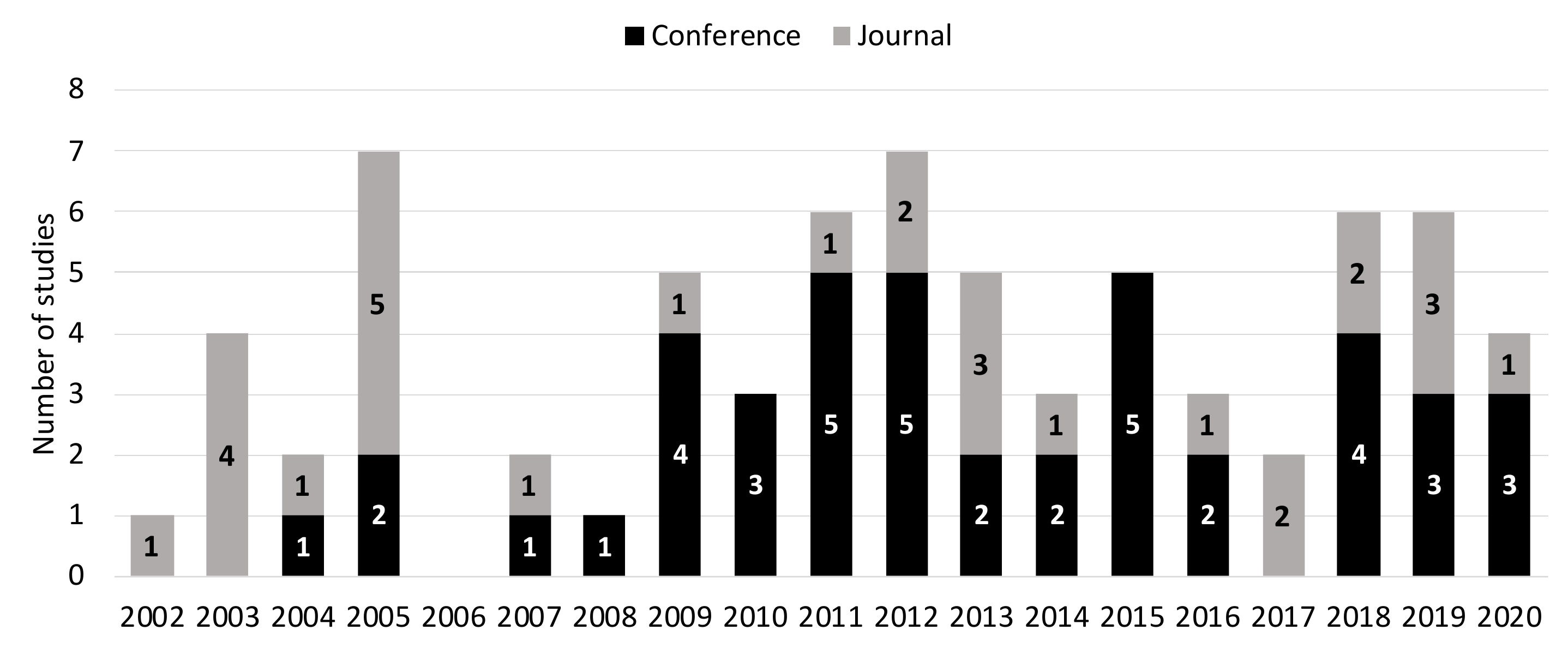}
        \caption{Distribution of studies over years and types of venues.}
        \label{fig:demographicinfo}
    \end{minipage}\hfill
    \begin{minipage}{0.49\textwidth}
        \centering
        \includegraphics[width=1\textwidth]{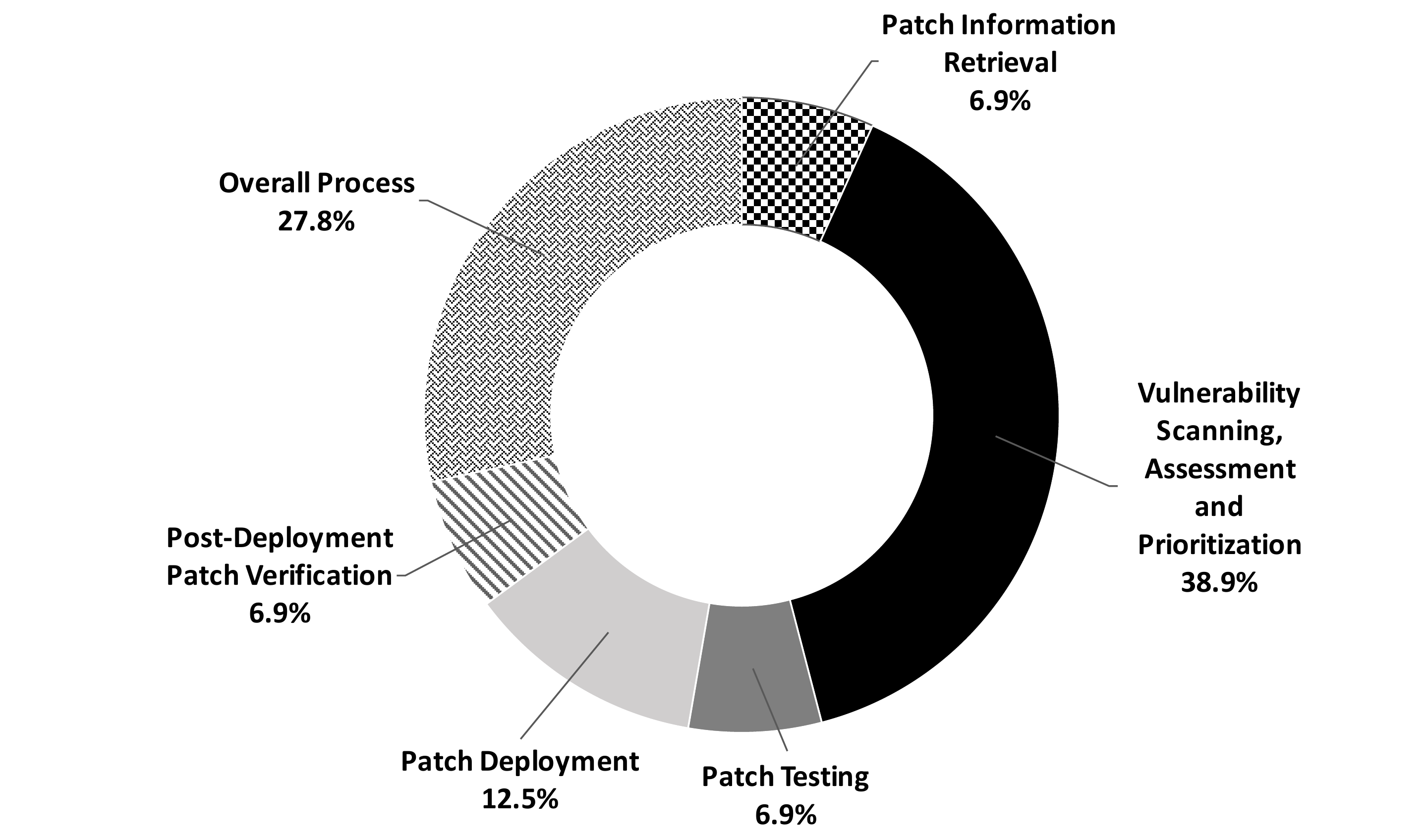}
        \caption{Distribution of studies over the software security patch management process.}
        \label{fig:focusarea}
    \end{minipage}
\end{figure}

Reporting demographic information in an SLR is considered useful for new researchers in that domain \cite{shahin2017continuous}. We present the demographic data of the distribution of the year and types of venues of the reviewed studies. Figure \ref{fig:demographicinfo} presents the distribution of 72 primary studies over the years and the different types of venues. The selected studies were published from 2002 to 2020 as we did not find any relevant studies published before 2002. We found that 60\% of the studies were published in conferences (43 of 72), while only 40\% studies appeared in scientific journals.


\subsubsection{Studies distribution in the software security patch management process}

We looked at the distribution of the reviewed studies mapped onto the software security patch management process discussed in Section \ref{section:background}. Figure \ref{fig:focusarea} reveals that a majority of the studies (38.9\%) focus on \emph{vulnerability scanning, assessment and prioritisation}. \emph{Patch information retrieval}, \emph{patch testing} and \emph{post-deployment patch verification} have received the least attention from the reviewed studies, with only 5 of 72 studies (6.9\%) focusing on those particular phases of the process. Twenty studies (27.8\%) focus on more than one phase of the process, which we classified under \emph{overall process}.


\subsubsection{Research Type}

\begin{figure}[!h]
    \centering
    \begin{minipage}{0.75\textwidth}
    \centering
      \includegraphics[width=0.9\linewidth]{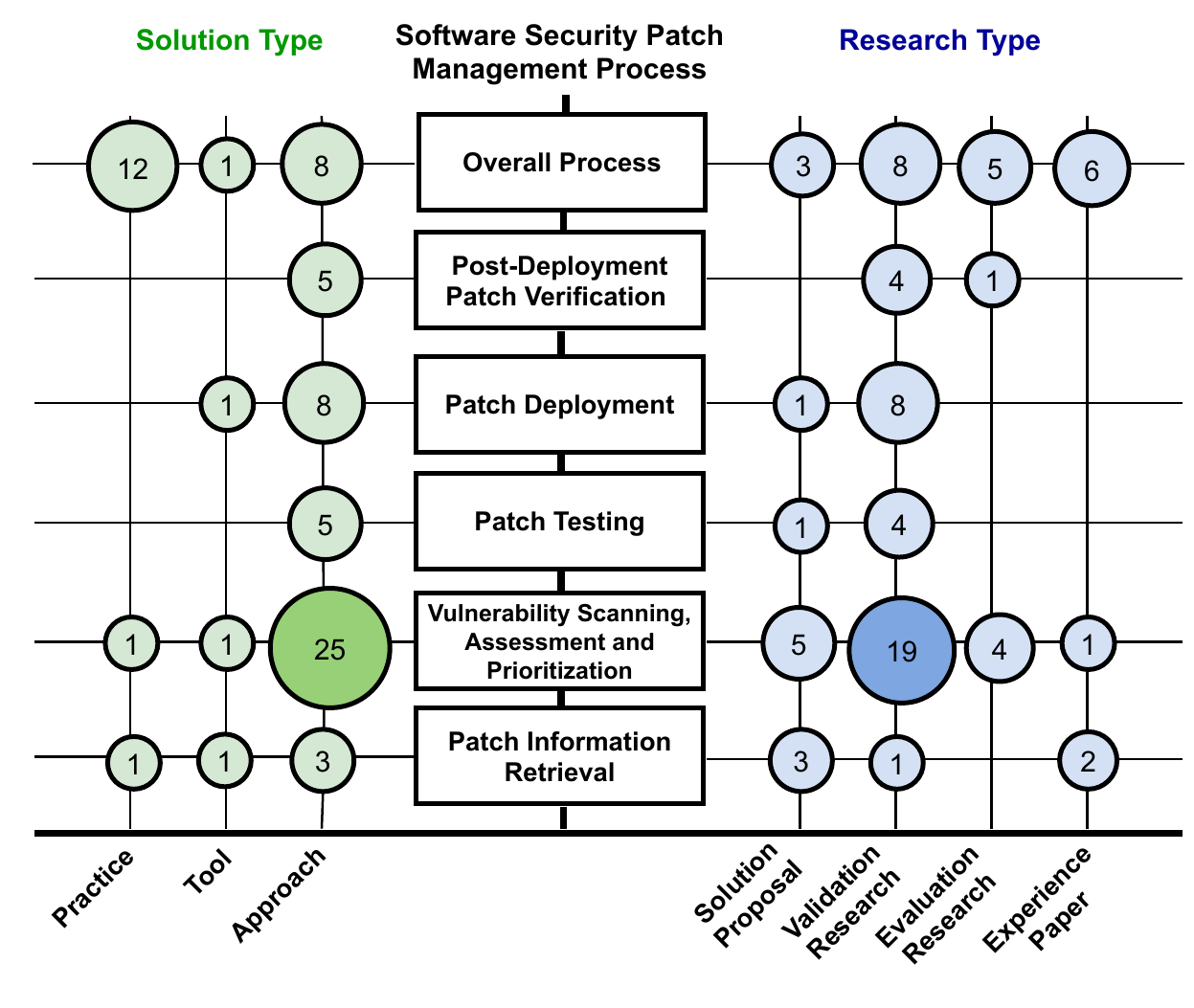}
      \captionof{figure}{Mapping of the research types and solution types with the software security patch management process.}
      \label{fig:researchcontrtypes}
    \end{minipage}
\end{figure}

We analysed the reported studies' research type and classified them into 4 categories, as illustrated in Figure \ref{fig:researchcontrtypes}, based on the classification proposed by Petersen et al. \cite{petersen2008systematic}. A majority of the studies (44, 61.1\%) reported \textit{validation research}, in which the dominant research methods consisted of simulation, laboratory experiments, mathematical analysis and prototyping \cite{petersen2015guidelines}. Thirteen studies (18.1\%) reported \textit{solution proposal} while only 10 studies (13.9\%) reported \textit{evaluation research} which consisted of strong empirical research methods such as industrial case study, controlled experiment with practitioners, practitioner targeted survey and interview \cite{petersen2015guidelines}. The least reported were 9 \textit{experience papers} (12.5\%) that included industrial experience reports. We did not find any studies related to the \textit{philosophical paper} and \textit{opinion paper} categories. Lack of evaluation research and experience papers indicate a large need for research aligned with industrial relevance in software security patch management.

\section{\textbf{Socio-technical challenges in software security patch management}}
\label{section:challenges}

This section presents the findings for RQ1, the socio-technical challenges in software security patch management. Our analysis resulted in the identification of 14 challenges as shown in Table \ref{tab:challenges}. We have classified the challenges that are common across all phases of the software security patch management process as ``common challenges" and others as specific to each phase of the process.

{\footnotesize\sffamily
\setlength\tabcolsep{2pt}
\begin{tabularx}{\textwidth}{>{\hsize=0.6\hsize}L
                                  >{\hsize=1.2\hsize}L
                                  >{\hsize=2.1\hsize}L
                                  >{\hsize=0.16\hsize}L}
\caption{Socio-technical challenges in software security patch management.}
\label{tab:challenges}\\
    \toprule 
\textbf{Relevant Patch Management Phase}  & \textbf{Challenges} & \textbf{Key Points (Included Papers)} & \textbf{\#} \\
    \midrule
\endfirsthead
   \toprule 
\textbf{Relevant Patch Management Phase}  & \textbf{Challenges} & \textbf{Key Points (Included Papers)} & \textbf{\#} \\
    \midrule
\endhead
    \bottomrule
\endfoot
    Common \newline{} Challenges & \textbf{Ch1:} Collaboration, coordination \newline{} and communication challenges & $\bullet$ Administrative overhead of coordinating with several stakeholders of conflicting interests [P4, P8, P11, P12, P30, P37, P63, P71]\newline{} $\bullet$ Delegation issues due to lack of accountability and well-defined roles and responsibilities [P20, P29, P30, P40]\newline{} $\bullet$ Communication challenges with multiple stakeholders with conflict of interests [P10, P15, P30, P34]\newline{} $\bullet$ Lack of collaboration among several stakeholders [P20] & 14   \\
    \cmidrule{2-4}
    &   \textbf{Ch2:} Impact of organisational policies/compliance & $\bullet$ Need to balance between complying with organisational policies and enforcing security [P8, P10, P20, P30, P71] & 5   \\
    \cmidrule{2-4}
    &  \textbf{Ch3:} Complexity of patches & $\bullet$ Diversity of patches (heterogeneity) [P15, P46, P49]\newline{} $\bullet$ Increasing rate of patch release [P1, P10, P29, P33, P35, P36, P40, P41, P44, P46, P55, P57, P62, P63, P66, P71]\newline{} $\bullet$ Large attack surface (large and distributed organisation structure) [P38, P42, P43] & 21   \\
    & \textbf{Ch4:} Limitations of existing tools     & $\bullet$ Lack of standardisation in heterogeneous tools [P10, P11, P18, P19, P25, P53, P64, P66, P71]\newline{} $\bullet$ Cost [P11, P19, P24, P25, P28, P36]\newline{} $\bullet$ Time-consuming [P23, P29, P36, P43]\newline{} $\bullet$ Lack of accuracy [P44, P45, P49, P51, P53, P67, P68, P69, P70]\newline{} $\bullet$ Lack of security [P1, P24, P30, P46, P47, P55, P56, P72]\newline{} $\bullet$ Lack of usability [P1, P25, P38, P45, P58, P70]\newline{} $\bullet$ Lack of scalability [P40, P54, P56, P58, P66, P68, P72]    & 33   \\
    \cmidrule{2-4}
    & \textbf{Ch5:} Need of human expertise & $\bullet$ Difficulty to achieve full automation in the process [P1, P9, P10, P11, P14, P16, P22, P25, P35, P48, P60, P62, P69, P70, P71]   & 15   \\
    \cmidrule{2-4}
    &   \textbf{Ch6:} Lack of resources & $\bullet$ Lack of skills and expertise [P1, P10, P15, P30, P45, P50]\newline{} $\bullet$ Lack of process guidelines [P2, P10, P14, P15, P20, P40, P69, P70]\newline{} $\bullet$ Lack of process automation solutions [P10, P19, P25] & 14   \\
    \addlinespace
    \hline
    \addlinespace
    Patch \newline{} Information \newline{} Retrieval   
    &   \textbf{Ch7:} Lack of a central platform \newline{} for information retrieval \newline{} and filtering     & $\bullet$ Lack of a unified platform for information retrieval [P8, P10, P15, P18, P53, P56, P65, P71]\newline{} $\bullet$ Lack of automatic validation, filtering and classification according to organisational needs [P8, P15, P18, P53, P56, P71]     & 8   \\
    \addlinespace
    \hline
    \addlinespace
    Vulnerability \newline{} Scanning, \newline{} Assessment and \newline{} prioritisation   
    &   \textbf{Ch8:} Lack of a complete \newline{} scanning solution     & $\bullet$ Lack of understanding of the system [P22, P26, P29, P37, P42, P48, P64, P65, P70]\newline{} $\bullet$ Lack of support for configuration management (detection) [P29, P32]\newline{} $\bullet$ Lack of knowledge of system inventories [P20, P48, P65]    & 10   \\ 
    \cmidrule{2-4}
     &   \textbf{Ch9:} Lack of support for \newline{} dynamic environment context     & $\bullet$ Inability to capture dynamic context-specific factors [P7, P13, P14, P32, P34, P41, P42, P49, P51, P55, P62]\newline{} $\bullet$ Lack of unified powerful metrics that capture the contextual factors [P6, P33, P51]     & 13   \\
    \cmidrule{2-4}
    &   \textbf{Ch10:} The gap of knowledge of technical and business context     & $\bullet$ Lack of knowledge of organisational business risk posture \newline{} and technical risk [P34, P50, P52, P71]     & 4   \\
    \addlinespace
    \hline
    \addlinespace
    Patch \newline{} Testing
    &   \textbf{Ch11:} Lack of proper \newline{} automated test strategy     & $\bullet$ Need for fully automated patch testing [P3, P8, P24, P59, P60, P68]     & 6   \\
    &   \textbf{Ch12:}  Poor test quality in \newline{} manual testing techniques     & $\bullet$ Difficulty dealing with patch dependencies [P4, P9, P10, P15, P16, P20, P60, P67, P68, P71, P72]\newline{} $\bullet$ Manual testing is slow and delays the patch deployment [P8, P12, P59, P71]\newline{} $\bullet$ Error-prone due to difficulty in exact replication of production state [P15, P29, P59, P60, P67]     & 15   \\ 
    \addlinespace
    \hline
    \addlinespace
    Patch \newline{} Deployment
    &   \textbf{Ch13:} Failures and side effects \newline{} due to deployment of patches     & $\bullet$ Need for managing the risk of problematic patches, missing configuration, and timely deployment [P3, P8, P9, P10, P12, P15, P16, P37, P45, P59, P60, P65, P71]\newline{} $\bullet$  Difficulty dealing with organisation constraints (system downtime) [P8, P10, P12, P17, P22, P25, P26, P30, P31, P37, P54, P58, P62, P63, P65, P66, P67, P71, P72]      & 27   \\ 
    \hline
    \addlinespace
    Post-Deployment \newline{} Patch \newline{} Verification
    &   \textbf{Ch14:} Lack of efficient \newline{} automated post-deployment \newline{} patch verification strategy    & $\bullet$ Lack of an overview of patch state of the system [P24, P37, P48]\newline{} $\bullet$ Issues with manual patch deployment verification: difficult, error-prone, time-consuming task [P5, P8, P12, P21, P48, P57]     & 8   \\ 
\end{tabularx}}

\subsection{\textbf{\underline{Common challenges}}}

The software security patch management process is a collaborative effort between multiple stakeholders such as internal teams including security managers, engineers and administrators, third-party software vendors like Microsoft, Adobe, and customers/end users. The conflicting interests and the interdependencies between these parties (e.g., delays of patch release from the third-party software vendors) make software security patch management a challenging undertaking [P4, P20, P30, P37]. The \textit{lack of collaboration, coordination and communication} between the involved stakeholders thus represents one of the main barriers in maintaining the security of the managed software systems [P4, P8, P11, P12, P30, P37, P63, P71]. Moreover, the \textit{impact of organisational policies}, i.e., the need to balance between complying with heterogeneous organisational policies and maintaining software security is acknowledged as a key challenge in software security patch management [P8, P10, P20, P30, P71]. This is because the policies set by the higher management (e.g., the minimum service interruptions policy) sometimes contradicts the timely application of emergency security patches [P20].

The rapid increase in the number and diversity of attacks has resulted in an increased rate of patch release causing a nightmare situation for practitioners to handle the increasing \textit{complexity of patches} [P1, P33, P35, P40, P41, P44, P46, P55]. Another contributor to the increased complexity of patches is the large and distributed attack vectors (i.e., the use of diverse software systems and products) in organisational environments [P38, P42, P43]. In addition, the \textit{limitations of the existing tools} have been noted as a major hindrance to the achievement of goals of software security patch management. Of them, some prominent limitations include the lack of a standard platform to integrate the heterogeneous tools used for patch management [P10, P11, P18, P19, P25, P53, P64, P66, P71], the lack of accuracy (e.g., the current tools failing to take into consideration the dynamic organisation context resulting in erroneous output [P44, P45, P49, P51, P53, P67, P68, P69, P70]), the lack of security [P1, P24, P30, P46, P47, P55, P56, P72], and the lack of scalability in the design/architecture of tools that create difficulties in applying patches to multiple systems with different operating systems.

Due to the increased complexity and dynamic nature of software security patch management and the limitations of the current technologies used in patching, the \textit{need for human expertise} is inevitable throughout the patching process [P1, P9, P10, P11, P14, P16, P22, P25, P35, P48, P60, P62, P69, P70, P71]. However, as a result of human involvement in patching tasks and decisions, the time to patch has increased leaving several attack vectors open to exploits [P1]. The risk of delays is further increased due to a \textit{lack of resources} in terms of skills and knowledge expertise, process guidelines and process automation support. An important point highlighted in the literature [P4, P29] regarding the lack of process automation support is that most of the existing solutions only focus on patch deployment, but do not provide solutions covering the entire process. Furthermore, several studies [P1, P19, P25, P30, P45, P50] have reported a significant gap in the required skills and knowledge expertise in software security patch management particularly due to the increased complexity of patches. 


\subsection{\textbf{\underline{Patch information retrieval related challenges}}} 


Practitioners are forced to spend hours monitoring multiple information sources due to the \textit{lack of a central platform for patch information retrieval and filtering} [P18, P53, P56]. Li et al. [P8] reported that modern patch information sources range from security advisories (78\%), official vendor notifications (71\%), mailing lists (53\%), online forums (52\%), news (39\%), blogs (38\%) to social media (18\%). Further, due to the rapid rate of patch releases, the lack of automated validation, filtering and classification of patch information according to organisational needs [P18, P53, P56] result in delayed patch application and increases the risk of a zero-day attack [P53].

\subsection{\textbf{\underline{Vulnerability scanning, assessment and prioritisation related challenges}}}

One of the prominent factors for the increased exposure to malicious attacks is the \textit{lack of a complete scanning solution}. As a result, the practitioners fail to obtain a clear understanding of the system leading to missing the detection of software vulnerabilities [P22, P26, P29, P37, P42, P48, P64, P65, P70] and system misconfigurations [P29, P32]. Concerning the vulnerability assessment and prioritisation, the \textit{lack of support for dynamic environment setting} presents a prominent challenge [P7, P13, P14, P32, P34, P41, P42, P49, P51, P55, P62]. The existing approaches are generally ``one size fits all'' that create difficulties in incorporating the needs of the organisational context and require a significant manual effort, particularly in virtual environment patching [P41, P42, P49, P51, P55]. A few studies [P6, P33, P51] have mentioned the need to have a common set of rigorous metrics with information such as exploit dates for accurate patch prioritisation since the existing vulnerability scanners depend on public vulnerability information which mostly includes only vulnerability disclosure dates. In addition, the \textit{knowledge gap of technical and business context} (e.g., the need to apply security patches as soon as possible vs prioritisation of systems' availability) often result in patch prioritisation conflicts between different teams [P34, P50, P52, P71].

\subsection{\textbf{\underline{Patch testing related challenges}}}

One of the most pressing challenges in modern patch testing can be attributed to the \textit{lack of a proper automated test strategy} [P3, P8, P24, P59, P60, P68]. A lack of automated testing may stem from different reasons such as the difficulty in dealing with issues of patch dependencies [P4, P9, P10, P15, P16, P20, P60, P67, P68, P71, P72] and the significant amount of human effort required for configuring a test environment to simulate a production-identical environment [P8, P59, P60]. However, although interestingly, most of the current patch testing is done manually to avoid the risks of unexpected system breakdowns caused by faulty and malicious patches [P8, P59, P60, P68]. Nevertheless, the \textit{poor test quality in manual patch testing} increases a system's vulnerability exposure as it often delays subsequent patch deployment [P8, P12, P59, P71]. Moreover, manual patch testing is largely error-prone due to the difficulty in exact replication of a production state [P15, P29, P59, P60, P67].

\subsection{\textbf{\underline{Patch deployment related challenges}}}

One of the central challenges facing modern organisations in software security patch management is the \textit{failures and side effects from the deployment of patches}. This challenge emerges as a result of poor patch testing leading to faulty patches being deployed and missing requisites for deployment such as the configuration and dependency changes causing deployment errors [P3, P8, P9, P10, P12, P15, P16, P37, P45, P59, P60, P65, P71]. Such errors would subsequently lead to additional service downtime, hence, many practitioners often delay or refuse to install patches and keep using outdated software instead leaving known vulnerabilities readily exploitable [P3, P37, P59, P60]. The other major challenge relates to dealing with the organisational constraints about system downtime. The lack of a proper run-time patch deployment strategy coupled with the organisational policies to avoid system downtime presents a serious challenge for timely patch installation [P8, P10, P12, P17, P22, P25, P26, P30, P31, P37, P54, P58,  P62, P63, P65, P66, P67, P71, P72]. This is particularly challenging in critical infrastructure system contexts such as healthcare for which downtime can create a significant adverse impact [P25, P30, P31, P37, P54, P58].

\subsection{\textbf{\underline{Post-deployment patch verification related challenges}}}

Most of the existing software security patch management solutions \textit{lack an efficient automated post-deployment patch verification strategy} offering an overview of the system's patch state. This results in difficulties for detecting the problem location when an issue occurs following patch deployment [P24, P37, P48]. In addition, most current patch auditing methods are manual requiring practitioners to manually inspect the application for signs of an attack and repair the damage if an attack is found. This is a frustratingly difficult and time-consuming task with no guarantee in finding every intrusion and reverting all changes exploited by an attacker [P21, P48, P57]. The need for this verification to be done as quickly as the patch is deployed adds to the complex, effort-intensive and time consuming manual verification highlighting the challenges of a lack of an efficient automated verification strategy [P5, P8, P12, P21, P48, P57].



\section{Approaches and tools proposed to facilitate software security patch management}
\label{section:approachesandtools}

This section presents the findings for RQ2.1, the approaches and tools proposed to facilitate the
software security patch management process. The analysis of the solution types has revealed that 75\% of the proposed solutions are \textit{approaches} while only 5.6\% of the solutions are \textit{tools}. Table \ref{tab:solareas} summarises the results for RQ2.1 presenting an overview of the key solution areas and the associated capabilities of the proposed approaches and tools, mapped onto the software security patch management process. 

{\footnotesize\sffamily
\setlength\tabcolsep{2pt}
\centering
\begin{tabularx}{\textwidth}{@{} >{\hsize=0.65\hsize}L
                                  >{\hsize=1.15\hsize}L
                                  >{\hsize=2.2\hsize}L
                                  >{\hsize=0.16\hsize}L}
\caption{A classification of solution areas and the associated capabilities of the reported approaches and tools.}
\label{tab:solareas}\\
    \toprule
\textbf{Relevant Patch Management Phase}  & \textbf{Solution Areas} & \textbf{Associated Capabilities (Included Papers)} & \textbf{\#} \\
    \midrule
\endfirsthead
   \toprule
\textbf{Relevant Patch Management Phase}  & \textbf{Solution Areas} & \textbf{Associated Capabilities (Included Papers)} & \textbf{\#} \\
    \midrule
\endhead
    \bottomrule
\endfoot
Patch \newline{}Information \newline{}Retrieval & \textbf{S1:} Patch Information Management & $\bullet$ Patch information retrieval from multiple sources (P18, P25, P28, P53, P56)\newline{} $\bullet$ Information filtering based on organisational configuration needs (P18, P56)\newline{} $\bullet$ Patch information validation (P18, P53, P56)\newline{} $\bullet$ Patch download and distribution (P105, P56, P28) & 5   \\
\addlinespace
\hline
\addlinespace
Vulnerability \newline{} Scanning, \newline{} Assessment and prioritisation & \textbf{S2:} Scanning for system vulnerabilities, potential attacks and ongoing attacks & $\bullet$ Central platform integrating the scan results from multiple sources (P22, P37, P41, P48, P51, P52, P64)\newline{} $\bullet$ Detailed host-based analysis to identify assets resident on host (P48, P51, P42)\newline{} $\bullet$ Detection of system misconfigurations (P32)\newline{} $\bullet$ Guidance on scanning tool selection (P69, P70)\newline{} $\bullet$ Identifying ongoing attacks (P22, P44)\newline{} $\bullet$ Providing historical scanning analysis (P22, P34, P42) & 27   \\
\cmidrule{2-3}
 & \textbf{S3:} Assessment and prioritisation of system vulnerabilities, \newline{} potential attacks \newline{} and ongoing attacks & $\bullet$ Providing a customisable, detailed and comprehensive analysis of vulnerability risks (P13, P26, P32, P34, P41, P43, P45, P49, P51, P52, P55)\newline{} $\bullet$ Prediction of optimal fixing strategy for potential and ongoing attacks (P22, P26, P43, P55)\newline{} $\bullet$ Measuring organisational vulnerability remediation effectiveness (P38, P43, P71)\newline{} $\bullet$ Capturing the dynamic context for accurate assessment and prioritisation (P7, P14, P23, P32, P38, P41, P44, P49, P51, P52, P55, P62) &    \\
Patch \newline{}Testing & \textbf{S4:} Automated detection \newline{} and recovery from faulty \newline{} and malicious patches & $\bullet$ Automated detection of faulty patches (P9, P16, P59, P60, P68)\newline{} $\bullet$ Automated detection of malicious patches (P24, P46, P47, P72) \newline{} $\bullet$ Automated crash recovery of faulty patches (P3) & 10   \\
\addlinespace
\hline
\addlinespace
Patch Deployment & \textbf{S5:} Automated patch deployment & $\bullet$ Consideration of the dynamic context in patch deployment (P9, P11, P16, P27, P31, P36)\newline{} $\bullet$ Reducing system downtime in reboots (P12, P17, P54, P58, P63, P67) & 15   \\
\addlinespace
\hline
\addlinespace
Post-Deployment Patch Verification & \textbf{S6:} Automated patch monitoring and \newline{} patch auditing & $\bullet$ Automated detection of exploits and patch deployment verification (P5, P9, P16, P24, P30, P37, P39, P48, P57, P72)\newline{} $\bullet$ Automated repair of past exploits (P21, P36) & 12   \\
\end{tabularx}}

\subsection{\textbf{\underline{Patch information retrieval related solutions}}}
    

A unified \textit{patch information management} platform including the capabilities of patch information retrieval from multiple sources, information filtering, classification, validation, download and distribution of patches benefits practitioners with timely patch information retrieval to protect from zero-day vulnerability attacks [P53]. Such a platform reduces the administrative overhead of having to monitor multiple information sources for receiving up to date patch information while providing an easy way to obtain patch information with high accuracy [P18, P25, P28, P53, P56]. However, verifying the information is important because the information is obtained through various sources that may contain non-validated information, for example, Twitter [P18]. To achieve that, Trabelsi et al. [P53] report a trust and reputation system to verify patch information using the KPI trust model.

\subsection{\textbf{\underline{Vulnerability scanning, assessment and prioritisation related solutions}}}

One of the first and foremost steps in securing software systems is \textit{scanning the systems to identify existing vulnerabilities, potential and ongoing attacks}. A set of studies have aimed at improving vulnerability scanning in different aspects. For example, a central platform that aggregates the scan results has been proposed to provide an overview of systems' patch state [P22, P37, P41, P48, P51, P52, P64]. This would serve as a proactive environment facilitating the identification of vulnerabilities, potential and ongoing attacks to assist practitioners with decision-making on the possible mitigation actions (e.g., applying patches, changing firewall rules, closing IP ports, etc.) [P22]. In an attempt to guide vulnerability scanning tool selection, Holm et al. [P69] find that there are significant differences in the accuracy of the scans of Windows and Linux hosts, through a comparative evaluation of seven tools used in the industry. 

Following the identification of vulnerabilities, potential and ongoing attacks in managed systems, \textit{accurate risk assessment} is essential for prioritising critical patches to protect against attacks. Approaches of a customised and comprehensive analysis of vulnerability risks have been proposed in line with the industry standard, the Common Vulnerability Scoring System (CVSS) \cite{CVSS} using different vulnerability characteristics [P13, P26, P32, P34, P41, P43, P45, P49, P51, P52, P55]. In addition, new quantitative metrics have also been introduced to consider the context-specific risks [P51, P52], for example, patch time and patch discovery time to consider the risk window between patch deployment and vulnerability scanning [P51]. Measuring the patch impact and effectiveness of remediation actions are equally important as devising timely remediation strategies for strategic planning. For that, solutions with real-time feedback on the remediation delays and analysis of patch applicability have been proposed [P38, P43, P71]. Several attempts [P7, P14, P23, P32, P38, P41, P44, P49, P51, P52, P55, P62] have been made to address the challenge of a lack of support for dynamic environment settings in risk assessment, particularly in the cloud, as the standard CVSS algorithm does not take into account the cloud-specific settings [P23, P49]. For example, the solution by Lin et al. is to consider temporal and environmental metrics on top of the base score in the current CVSS algorithm. 

\subsection{\textbf{\underline{Patch testing related solutions}}}

The necessity for rigorous patch testing emerges from the existence of faulty and malicious patches. To overcome the identified challenges of rapid patch release rates and poor test quality in manual patch testing techniques, approaches for \textit{automated detection and recovery from such faulty and malicious patches} have been proposed [P9, P16, P59, P60, P68]. For example, Maurer and Brumley [P59] propose a tandem execution approach that immediately detects vulnerability exploits with no false positives. Few studies [P24, P72] have proposed using Blockchain to ensure the integrity of patches that are resilient to malicious attacks during patch distribution. Although several attempts have been made at detecting faulty and malicious patches, we found only one solution [P3] for surviving crashes that result from faulty patches. The proposed solution is based on multi-version execution and helps achieve minimal disruption to operations during a crash.

\subsection{\textbf{\underline{Patch deployment related solutions}}}

Several solutions have been proposed for \textit{automating patch deployment} extending context-specific support for distributed and heterogeneous environments and reducing the patch deployment time, cost and overhead [P19, P25, P27, P31, P36, P40, P54, P58]. Additionally, reducing system downtime and reboots have been a priority of several studies [P12, P17, P54, P58, P63, P67] to address the critical challenge of minimising service interruptions when deploying patches. For example, Yamada et al. [P54] propose a virtual machine monitor (VMM)-based approach namely Shadow Reboot, to shorten the downtime and enable applications to run while rebooting. This approach can serve as a complementary solution to the existing dynamic software updating methods that usually require practitioners to have knowledge about the target kernels at the source code level [P54]. Alternatively, an approach that achieves minimal downtime through instant kernel updates without additional modifications to programs or state change tracking has been proposed by Kashyap et al. [P58]. They use an application checkpoint and restart (C/R) method to reduce the downtime to just three seconds.

\subsection{\textbf{\underline{Post-deployment patch verification related solutions}}}



Several solutions have been suggested for \textit{automating post-deployment patch verification tasks} [P5, P9, P16, P24, P30, P37, P39, P48, P57, P72] and \textit{repairing past exploits} [P21, P36]. For example, ``Pakiti" [P30] is a system that provides a central view of the patching status to help practitioners be informed and detect problems following patch deployment. Concerning the focus on automating the repair of past exploits, ``Nuwa" [P36] is a tool that can automatically detect and repair patch deployment failures. It allows practitioners to retroactively patch vulnerabilities by automatically repairing the changes that have resulted from exploits while maintaining legitimate user changes [P21].

\section{Practices proposed to successfully implement software security patch management}
\label{section:practices}

This section presents the findings for RQ2.2, the practices to successfully implement the software security patch management process. 
The classification of practices includes the reported recommendations, guidelines, best practices, lessons learned and shared experiences of researchers and industry practitioners. Similar to RQ1, we present the practices that are common to all phases of the software security patch management process and those that are specific to each phase as shown in Table \ref{tab:practices}.



{\footnotesize\sffamily
\setlength\tabcolsep{2pt}
\centering
\begin{tabularx}{\textwidth}{>{\hsize=0.5\hsize}L
                                  >{\hsize=1.3\hsize}L
                                  >{\hsize=2.1\hsize}L
                                  >{\hsize=0.15\hsize}L}
\caption{Practices proposed for successful implementation of software security patch management.}
\label{tab:practices}\\
    \toprule 
\textbf{Relevant Patch Management Phase}  & \textbf{Practices} & \textbf{Key Points (Included Papers)} & \textbf{\#} \\
    \midrule
\endfirsthead
   \toprule 
\textbf{Relevant Patch Management Phase}  & \textbf{Practices} & \textbf{Key Points (Included Papers)} & \textbf{\#} \\
    \midrule
\endhead
    \bottomrule
\endfoot
Common \newline{} Practices & \textbf{PR1:} Planning and documentation & $\bullet$ Document the life cycle of each vulnerability including reporting and tracking of remediation measures (P20, P66)\newline{} $\bullet$ Review and update the process on a bi-yearly basis (P20, P61)\newline{} $\bullet$ Time and dedication need to be given for proactive planning (P20) & 3   \\
\cmidrule{2-4}
& \textbf{PR2:} Establish formal policies and procedures into process activities & $\bullet$ Develop an appropriate mitigation strategy when no patch/workaround is supplied by the vendor (P20) \newline{} $\bullet$ Have formal processes defined into the process covering all phases of the process (P2, P20, P50, P61, P65, P66, P72)\newline{} $\bullet$ Measure the performance and effectiveness of the process (P2, P20)\newline{} $\bullet$ Formalise procedures for dispute resolution (P2, P20) & 8   \\
\cmidrule{2-4}
& \textbf{PR3:} Define roles and responsibilities in the process & $\bullet$ Define the roles and responsibilities of groups and individuals involved in the process (P2, P20, P40, P61, P72) \newline{} $\bullet$ Require stakeholders to take accountability (P20) & 5   \\
& \textbf{PR4:} Get management involvement and a clear understanding of the process & $\bullet$ Get senior management approval and involvement in process activities (P20, P40, P72)\newline{} $\bullet$ Require clear understanding of the process for accurate decision-making (P8, P20, P61) & 5   \\
\cmidrule{2-4}
& \textbf{PR5:} Define procedures to facilitate efficient communication and collaboration & $\bullet$ Establish procedures to enable efficient communication and collaboration between stakeholders (P20, P61, P66) \newline{} $\bullet$ Hold frequent patch meetings (P20)\newline{} $\bullet$ Increase stakeholders' awareness of the process (P50)\newline{} $\bullet$ Coordinate patch release schedules of different vendors (P4) & 5   \\
\addlinespace
\hline
\addlinespace
Patch \newline{} Information \newline{} Retrieval & \textbf{PR6:} Establish policies and responsibilities for information retrieval, notification and \newline{} dissemination & $\bullet$ Establish and maintain a list of the information resources (P2, P8, P19, P35, P65) \newline{} $\bullet$ Maintain an upstream and downstream infrastructure for patch download and distribution to limit latency (P28)\newline{} $\bullet$ Have proper patch information notification and dissemination policies in place (P35) & 6   \\
\addlinespace
\hline
\addlinespace
Vulnerability Scanning, Assessment \newline{} and \newline{} Prioritisation & \textbf{PR7:} Regularly monitor both \newline{} active and inactive applications \newline{} and security intelligence sources & $\bullet$ Regularly scan and monitor the network \newline{} and vulnerability alerts (P2, P4, P20, P34, P40) \newline{} $\bullet$ Establish a dedicated mailbox for vulnerability alerts \newline{} that are sent via email (P2)\newline{} $\bullet$ Close down unnecessary ports on network devices (P34)\newline{} $\bullet$ Maintain historical scanning reports for future analysis (P34) & 4   \\
\cmidrule{2-4}
& \textbf{PR8:} Maintain up to date system inventory & $\bullet$ Create and maintain a system inventory including \newline{}  all the previous patches installed on every system (P20, P40, P61, P72) \newline{} $\bullet$ Classify assets by platform hardware type, location and software application records, and develop risk potential \newline{} for each asset (P2, P65, P66) & 7   \\
\cmidrule{2-4}
& \textbf{PR9:} Perform vulnerability \newline{} assessment based on \newline{} organisation needs and context & $\bullet$ Organisations should perform their own vulnerability assessment (P50, P61, P65) \newline{} $\bullet$ Assess and respond to vulnerabilities on time (P2, P20, P50) \newline{} $\bullet$ Consider historical scanning analysis in risk assessment (P34) & 6   \\
\addlinespace
\hline
\addlinespace
Patch \newline{} Testing & \textbf{PR10:} Improve testing activity & $\bullet$ Prepare and store the test environment for manual \newline{} system testing (P19) \newline{} $\bullet$ Develop and test back-out procedure (P2, P20, P40) & 4   \\
Patch Deployment & \textbf{PR11:} Install patches on time balancing the security risks, \newline{} resources and system availability & $\bullet$ Install timely patches balancing the need for security, resources, and time required to test a patch for system stability (P2, P20, P35, P40, P50, P65, P66, P72) \newline{} $\bullet$ Facilitate automation as much as possible in the process (P35)\newline{} $\bullet$ Investigate workarounds to reduce system reboots (P35, P50, P66)\newline{} $\bullet$ Define a matrix for patch scheduling by patch severity and profile of managed systems (P19, P66) & 9   \\
\addlinespace
\hline
\addlinespace
Post-Deployment Patch Verification & \textbf{PR12:} Keep track of the deployment status of every patch  & $\bullet$ Regularly monitor system's patch status to make sure every single patching job is executed successfully (P19, P40, P65, P66) & 4   \\
\end{tabularx}}

\subsection{\textbf{\underline{Common Practices}}}

A \textit{well-planned and structured process} is vital for successful software security patch management. To define a solid patch management process, an organisation needs to give its time and dedication upfront [P20]. The process should \textit{establish formal policies and procedures in the process activities} in all phases including documentation, communication, management reporting and auditing [P2, P20, P50, P61, P65, P66, P72]. For example, standard procedures should be in place for dispute resolution to handle conflicts and escalation paths for emergencies [P2, P20]. It is also important to measure the performance and effectiveness of the defined policies and procedures in the process and update them accordingly on a bi-yearly basis [P2, P20]. 

Since several internal and external stakeholders are involved in the process, having \textit{well-defined roles and responsibilities} of individuals and groups helps reduce the administrative overhead of coordinating with multiple stakeholders and increases task accountability [P2, P20, P40, P61, P72]. According to Nicastro [P20], a local Computer Incident Response Team (CIRT) and Information Risk Managers (IRMs) are some of the roles that should be defined in every organisation. As such, adhering to a standard set of procedures, roles and responsibilities can help achieve \textit{a clear understanding of the process among all stakeholders} and minimise conflicts [P8, P20, P61]. It is also important to have the \textit{senior management actively involved and supporting the patch management decisions} to obtain organisational approvals without delays [P20, P40, P72]. Finally, \textit{efficient communication and collaboration} between all stakeholders are vital for smooth execution of the process [P20, P61, P66].
         


\subsection{\textbf{\underline{Patch information retrieval related practices}}}

    
\textit{Developing formal policies and responsibilities for patch information retrieval, patch download and dissemination} (e.g., creating and maintaining a list of information sources) has been reported as useful practices to reduce the latency in information retrieval-related activities [P2, P8, P19, P28, P35, P65].

\subsection{\textbf{\underline{Vulnerability scanning, assessment and prioritisation related practices}}}

Concerning vulnerability scanning, \textit{regularly monitoring both active and inactive applications and security intelligence sources} is important since there is a possibility of exploitation for applications that are not frequently used [P4]. Maintaining a history of the scanning reports [P34] is useful for analysing trends and making predictions of potential attack opportunities. Another best practice for vulnerability scanning is to \textit{maintain an up to date system inventory} to increase the understanding and awareness of the system. With regards to the vulnerability assessment, \textit{organisations should perform their own risk assessment based on the context} instead of solely relying on the vulnerability assessment scores from software vendors to get a more accurate vulnerability risk.

\subsection{\textbf{\underline{Patch testing related practices}}}


    

As identified in Section 4.4, most current patch testing is done manually to avoid the risks of unexpected system breakdowns from faulty and malicious patches. To \textit{improve testing activities}, it is proposed to prepare the test environment including pre-configuration tasks and storage in advance to save time in testing [P19]. Although some practitioners avoid testing small patches due to the large overhead involved with patch testing, the authors in [P2, P20, P40] highlight the necessity for testing all security patches, and developing and testing the back-out procedure to be deployed when required [P20].

\subsection{\textbf{\underline{Patch deployment related practices}}}

    

The \textit{patches need to be installed on time while balancing the risks of time for proper patch testing and potential attacks while effectively managing the organisation constraints } (e.g., service availability constraints). According to Marx et al. [P50], ``a successful patch management process is capable of patching vulnerabilities in the shortest possible time frame while preventing the system downtime caused by an insufficiently tested patch". The path to achieving this balance is to have an appropriate risk-focused patch management process [P50] and proper patch scheduling (e.g., defining a matrix for scheduling patches based on the patch severity and its impact on the managed systems) [P19, P66].

\subsection{\textbf{\underline{Post-deployment patch verification related practices}}}
    
    

\textit{Keeping track of the deployment status of every patch} is useful to verify the deployment of patches, detect post-deployment issues early, and ensure the potential exploits during patch deployment are properly identified and repaired. To achieve this, it is suggested to regularly monitor a system's patch status and seek client feedback for any adverse impact on service continuity after every patching job [P19, P40, P65, P66].
    
\section{Evaluation of the reported solutions in software security patch management}
\label{section:evaluation}

In this section, we report the results for RQ3, on how well the solutions have been assessed. We adopted the classification of evaluation approaches proposed by Chen et al. \cite{chen2011systematic} presented in Table \ref{tab:evalcategory} to categorise the evaluation types used in the reviewed studies. We have slightly modified the adopted classification with two additions (i.e., \textit{``SR - Simulation with real data"} and \textit{``NE - No Evaluation"}) to make it more suitable for our review.

{\footnotesize\sffamily
\setlength\tabcolsep{4pt}
\centering
\begin{tabularx}{\textwidth}{   >{\hsize=0.6\hsize}L
                                >{\hsize=1.2\hsize}L}
\caption{The scheme for classification of the evaluation types.}
\label{tab:evalcategory}\\
        \toprule
        \textbf{Evaluation type} & \textbf{Definition} \\
            \midrule
        \endfirsthead
           \toprule
        \textbf{Evaluation type} & \textbf{Definition} \\
            \midrule
        \endhead
            \bottomrule
        \endfoot
        Field experiment & Controlled experiment performed in industry settings \\
        \addlinespace
        Case study & An empirical study that investigates a contemporary phenomenon within \newline{} its real-life context; i.e., studies involving industry practitioners \cite{shaw2003writing} \\
        \addlinespace
        Experience & The result has been used on real examples, but not in the form of case studies \newline{} or controlled experiments, the evidence of its use is collected informally or \newline{} formally. e.g., industrial experience reports \\
        \addlinespace
        Simulation with artificial data & Execution of a system with artificial data, using a model of the real world \\
        \addlinespace
        Simulation with real data & Execution of a system with real data, using a model of the real world \newline{} performed in laboratory experiment \\
        \addlinespace
        Laboratory experiment with software subjects & A laboratory experiment to compare the performance of newly proposed \newline{} system with other existing systems \\
        \addlinespace
        Laboratory experiment with human subjects & Identification of relationships between variables in a designed controlled \newline{} environment using human subjects and quantitative techniques \\
        \addlinespace
        Rigorous analysis & Rigorous derivation and proof, suited for formal model (i.e., statistical \newline{} or mathematical verification) \\
        \addlinespace
        Discussion & Provided some qualitative, textual, opinion-oriented evaluation. e.g., compare \newline{} and contrast, oral discussion of advantages and disadvantages \\
        \addlinespace
        Example application & Authors describing an application and provide an example to assist in the \newline{} description for evaluation \\
        \addlinespace
        No Evaluation & A study that reports no evaluation \\
\end{tabularx}}

\subsection{\textbf{\underline{Types of evaluation used to assess the proposed solutions}}}

\begin{figure}[h!]
    \centering
    \begin{minipage}{1\textwidth}
    \centering
    \includegraphics[width=1\linewidth]{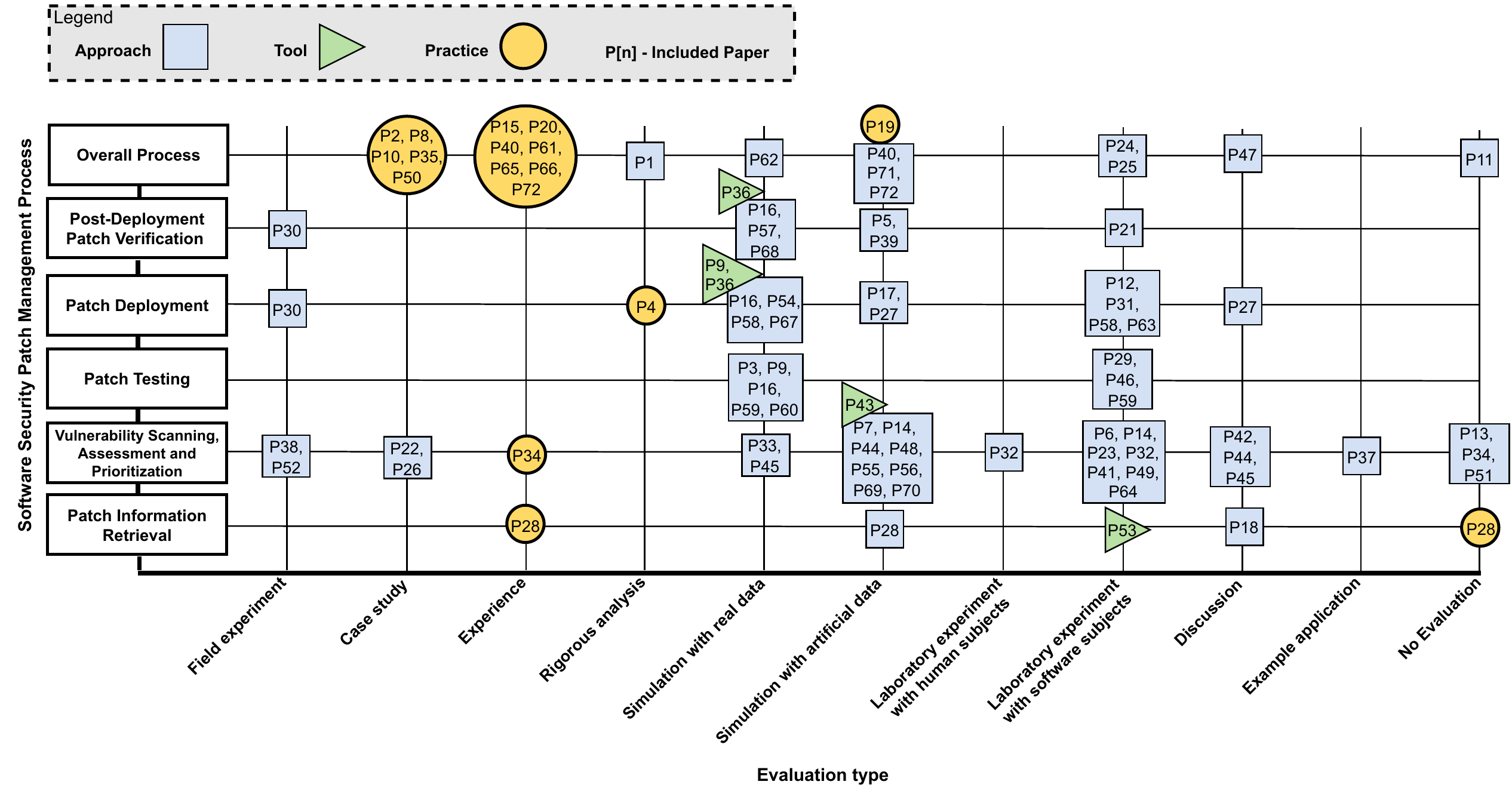}
    \captionof{figure}{Mapping of the evaluation types based on the solution type and software security patch management phase (symbol size based on the number of papers per solution type).}
    \label{fig:evalmapping}
    \end{minipage}
\end{figure}

The importance of rigorous evaluation to assess the appropriateness of the proposed solutions has been emphasised by the software engineering research community \cite{shaw2003writing, zannier2006success, chen2011systematic}. Accordingly, in this subsection, we present the results for RQ3.1, a distribution of the evaluation types used in the proposed solutions based on the solution type (i.e., approach/tool/practice) and the software security patch management phase, as shown in Figure \ref{fig:evalmapping}.  

With regards to the evaluation types of the proposed \textit{approaches}, \textit{``Laboratory experiment with software subjects"} (30.36\%) is the most frequently used type followed by \textit{``Simulation with artificial data"} (28.57\%). Interestingly, four studies [P11, P13, P34, P51] have reported \textit{``No evaluation"} of their proposed approaches while evaluation types such as \textit{``Case study"} (3.57\%), \textit{``Laboratory experiment with human subjects"} (1.79\%) and \textit{``Rigorous analysis"} (1.79\%) have been rarely used. Concerning the \textit{tools}, the majority (50\%) have been evaluated using \textit{``Simulation with real data"}. Regarding the \textit{practices}, \textit{``Experience"} has been a widely used method for assessment (56.25\%). Five studies (31.25\%) have used \textit{``Case study"} whereby the real-world insights have been captured through research methods such as industrial pilot projects [P2], practitioner targeted surveys [P8, P10, P35, P50] and interviews [P8, P10]. One study by Nappa et al. [P4] proposes a set of recommendations for patch deployment based on rigorous analysis of 1593 client-side vulnerabilities.

It should be noted that 11 of the reviewed studies (15.3\%) [P14, P27, P28, P32, P34, P40, P44, P45, P58, P59, P72] have used two types of evaluation to assess their proposed solutions. For example, Xiao et al. [P44] used \textit{``Simulation with artificial data"} to evaluate the robustness of the proposed approach against vulnerability exploits, and theoretical reasoning (i.e., \textit{``Discussion"}) to demonstrate its practical utility for real-world monitoring of software vulnerabilities. Similarly, some studies have proposed an \textit{approach} and a \textit{tool} that address the challenges across multiple patch management phases (e.g., [P9] presents a method and a tool to analyse the patch impact and support patch deployment).

\subsection{\textbf{\underline{The level of rigour and industrial relevance of the reported solutions}}}

The importance of providing practitioners with solutions to real problems and understanding how well the solutions have been evaluated cannot be overlooked in software engineering research \cite{fenton1994science}. Correspondingly, this subsection reports our attempts in assessing the level of rigour and industrial relevance of the evaluation types used for assessing the reported solutions as findings for RQ3.2. Of the evaluation types listed in Table \ref{tab:evalcategory}, \textit{``Field experiment"} is considered the most rigorous form of evaluation, followed by \textit{``Case study"}. It is because both methods have the highest industrial relevance as the evaluation involves industry practitioners \cite{petersen2015guidelines}. Similarly, evaluation based on \textit{``Experience"} (e.g., industrial experience reports) also results in industry-relevant outcomes. By contrast, all other evaluation types are not acknowledged as rigorous forms of evaluation with proper industrial relevance. It is considered that \textit{``Discussion"} and \textit{``Example application"} evaluation types contain the least rigour and industrial relevance.

Since software security patch management is a highly industry-oriented topic, employing evaluation types of \textit{``Field experiment"} and \textit{``Case study"} would produce solutions having higher industry adaptation and usefulness. However, a concerning finding is that only 15 solutions (20.8\%) have used an evaluation type with industrial relevance. Of those 15 studies, seven solutions have been evaluated using \textit{``Case study"}, three with \textit{``Field experiment"} and the remaining five solutions using \textit{``Experience"} (i.e., industrial experience reports). Another notable finding from this analysis is the lack of replication studies in the reviewed studies. According to Chen et al. \cite{chen2011systematic}, replication helps to provide solid and reliable evidence to support the adoption of a particular solution. We have found that 65 studies (90.3\%) have evaluated their solutions in only one study indicating a general lack of replication. These findings reveal that majority of the proposed solutions lack rigorous and industry suitable evaluation, which is alarming given that the domain is highly industry-centric.

\section{Discussion}
\label{section:discussion}

\begin{figure}[h!]
    \centering
    \begin{minipage}{1\textwidth}
    \centering
      \includegraphics[width=1\linewidth]{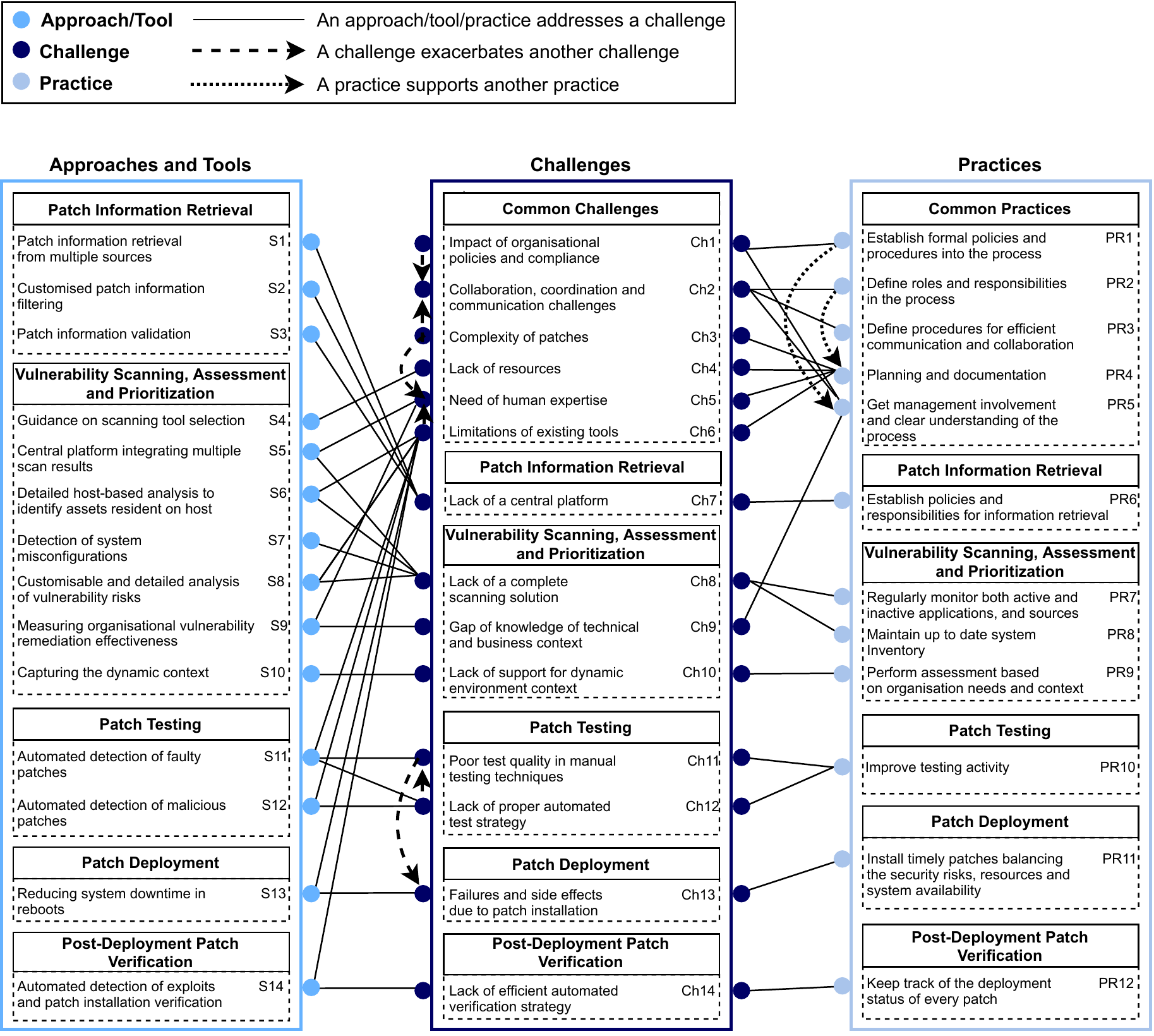}
      \captionof{figure}{A mapping of challenges onto solutions.}
      \label{fig:discussionmapping}
    \end{minipage}
\end{figure}

In this section, we discuss the key findings from this SLR and the potential future research and development opportunities in this domain based on the key limitations and gaps identified through our study findings. We present a mapping of our findings in Figure \ref{fig:discussionmapping} to enable the reader to quickly identify the relationships between the challenges (Section \ref{section:challenges}) and the proposed solutions (i.e., approaches and tools (Section \ref{section:approachesandtools}), and practices (Section \ref{section:practices})) and the dependencies between them. An important observation is the dependencies exist only among the challenges and practices, and that they can be classified into two types, \textit{dependencies among challenges that negatively affect or exacerbate another challenge} and \textit{dependencies among practices that positively affect or support another practice}, as illustrated in Figure \ref{fig:discussionmapping}. For example, the \emph{lack of a proper automated test strategy} exacerbates the issues with \emph{poor test quality in manual testing techniques}. Consequently, it leads to faulty patches deployed causing \emph{failures and side effects} during patch deployment. Alternatively, \emph{establishing formal policies and procedures into process activities} helps \emph{obtain a clear understanding of the process and obtain approval for software security patch management decisions from the management without delays}.

\subsection{Need for more investigation on the less explored software security patch management phases} 

As shown in Figure \ref{fig:focusarea}, out of the primary studies selected, 28 (38.9\%) have exclusively focused on proposing solutions to address the challenges in vulnerability scanning, assessment and prioritisation phase, 20 studies (27.8\%) have focused on facilitating more than one software security patch management process phase, and nine studies (12.5\%) have proposed solutions to patch deployment phase. However, an important realisation from the analysis is the lack of attention paid to patch information retrieval, patch testing and post-deployment patch verification phases where only five studies (6.9\%) have focused on each phase. The patch management process represents a tightly coupled sequence of phases where the output of one phase is input to the next phase. In addition, the dependencies among them exacerbate the challenges in the execution of tasks as identified in Figure \ref{fig:discussionmapping} (e.g., [Ch12, Ch11] $\underrightarrow{exacerbate}$ Ch13). Hence, there is a clear opportunity for valuable future research focusing on these less explored phases of the patch management process.

\subsection{Need for focus on socio-technical aspects in software software security patch management}

Software security patch management is predominantly a socio-technical phenomenon where the human and technological issues/solutions are intertwined demanding exhaustive collaboration and coordination between human-human and human-technical systems interactions \cite{dissanayake2021grounded}. The results in Section 4.1 have revealed that lack of efficient collaboration, coordination and communication [Ch2] during the software security patch management process can cause a significant negative influence on the timely vulnerability remediation. However, as shown in Figure \ref{fig:discussionmapping}, this important challenge has not received much attention in the reviewed studies. Based on our findings (see Figure \ref{fig:discussionmapping}), we argue that the patch interdependencies intrinsic to software security patches resulting in increased complexity of patches [Ch3] can have a significant effect on other socio-technical factors like collaboration and coordination [Ch2]. Subsequently, this may result in additional struggles for organisations to apply timely patches. This apparent influence of socio-technical challenges to the success of the software security patch management process implies a need for more research on the roles and effects of such socio-technical aspects in software security patch management. Moreover, we believe such findings would be useful for software developers to consider the socio-technical aspect integral to patch application when developing patches. Additionally, future studies can invest effort into developing tools and frameworks to support the socio-technical aspects, for example, a tool enabling better coordination across patch management tasks and multiple stakeholders.

\subsection{Human-AI collaboration for securing software systems}

Over the years, several attempts have been made to integrate automation into software security patch management tasks. However, an important realisation in the results presented in Section 4.1 is that there needs to be a delicate balance between human intervention and automation in software security patch management. Automation enables practitioners to enjoy the benefits of less manual effort, while human expertise is required in the loop taking control of the decision-making and tasks that cannot be completely automated due to complexities of patches [Ch3] and limitations in the technology [Ch6]. Our review has unveiled that such decision-making points exist throughout the software security patch management process. For example, according to P59, \emph{``in an aim to automate the patch testing as much as possible, it is noted that the human intervention is inevitable. As patches can change the semantics of a program, a human will likely always need to be in the loop to determine if the semantic changes are meaningful"}. As noted in Figure \ref{fig:discussionmapping}, only a few studies have addressed this challenge opening several possibilities for future research in \emph{``Human-AI Collaboration"} \cite{kamar2016directions}, which is a new and emerging research paradigm aimed at exploring how human intelligence can be integrated with an Artificial Intelligence (AI) systems to complement machine capabilities. Such systems aim at achieving complex tasks through hybrid intelligence to collectively achieve goals and progressively improve by learning from each other \cite{dellermann2019future}. We believe that an intelligent system with a real-time, human-like, cognition-based framework can guide practitioners to decisions by providing timely information and asking logical questions, thereby assisting autonomous decision-making in software security patch management \cite{crowder2020human}.

\subsection{Standardisation of heterogeneous tools}

As identified in section 4.1, the diversity (heterogeneity) of patches increases patch complexity [Ch3]. This usually results in several challenges to practitioners from patch information retrieval to post-deployment patch verification. In addition, organisations use a multitude of software products (e.g., operating systems (OS), software applications, tools and platforms) increasing the challenges of patch heterogeneity. Several studies [P40, P54, P56, P58, P66, P68, P72] have reported the limitation of the lack of scalability in the existing tools. It was also observed that the majority of the reported solutions are only compatible with Linux, possibly due to the reasons of its being open source, easier to configure than other OSs and that patches applied to many Linux distributions result in only minor changes as opposed to patches of Windows [P69]. Hence, there is an increasing realisation of an orchestrated platform that caters to these heterogeneous tools. Future research could focus on designing and evaluating an architecture to support the standardisation of heterogeneous patch management tools which is dynamically adaptable to the organisational context and needs.

\subsection{Real-world, rigorous evaluations}

As revealed by the results in Section \ref{section:evaluation}, a large number of studies lack rigorous evaluation using more mature forms of evaluation like field experiments and case studies. 
The low percentage of the studies with industry-related evaluation highlights the need for researchers to work with practitioners to improve the state of the practice of rigorously evaluating research outcomes. We recommend that more attention be paid to rigorously evaluate the solutions using approaches with industrial relevance. The robust evaluation will improve the quality and transferability of the research outcomes to industrial adoption.

\subsection{Contextual factor}

The importance of reporting contextual factors have been emphasised in the literature stating that software engineering research should investigate and understand their respective context \cite{kirk2014investigating, dybaa2012works, petersen2009context}. In our review, we have tried to identify how the studies have reported the methodological and organisational context (i.e., research type, solution type). It should be noted that some studies had to be included in two categories since they have reported two research types. 
Based on the findings in Section 3.7.3 and Section \ref{section:evaluation}, only 20.8\% of the reviewed studies have provided industry-related evidence. 
The industry-related evidence enables researchers to understand the practical utility of the reported solutions and practitioners to adopt the proposed solutions in the literature. Given software security patch management is a very industry-centric topic and complementing the need highlighted by previous literature [e.g., P10], we recommend future research focus more attention on reporting the contextual factors as it helps to increase the credibility and quality of the research.

\section{Threats to Validity}
\label{section:threats}

In this section, we report the validity threats of our study and the corresponding mitigation strategies following the guidelines proposed by \cite{ampatzoglou2019identifying, Kitchenham07guidelinesfor, zhou2016map}.

\subsection{Internal validity} 
Bias in study selection (i.e., study filtering) and data extraction represent standard threats to all SLRs \cite{zhou2016map}. To address this, we defined a review protocol with explicit details about the search string construction, search process, study inclusion/exclusion and data extraction strategy \cite{ampatzoglou2019identifying, dybaa2008empirical, unterkalmsteiner2011evaluation}. Following a well-defined protocol helps achieve consistency in the study selection and data extraction, particularly, if multiple researchers are involved in the process \cite{unterkalmsteiner2011evaluation}. We iteratively developed and improved the protocol, particularly the inclusion/exclusion criteria, after conducting a staged study selection process and pilot data extraction. Further, two authors selected the studies while the other authors cross-checked the outcomes and appropriateness of the selection criteria using randomly selected papers. 

Concerning data extraction bias, we executed a pilot data extraction on a randomly selected sample of five studies to ensure the data extraction form captures all the required data to answer the RQs. We used a data extraction form (adapted from \cite{keele2007guidelines, garousi2019guidelines}) which was reviewed by all authors through the pilot data extraction. The first author extracted the data which was cross-checked by the other authors for accuracy. Throughout study selection and data extraction phases, weekly detailed discussions were held between all authors to resolve the disagreements.

Additionally, publication bias is acknowledged as an internal validity threat which refers to the issue of the high likelihood of publishing positive results than negative ones \cite{keele2007guidelines}. However, we have reported the negative results captured in the primary studies (e.g., challenges in software security patch management (RQ1)) and the challenges have been mapped against the reported solutions (RQ2), i.e., the positive results when identifying the gaps in Section \ref{section:discussion} moderating the effect of unreported negative results. Further, using snowballing to increase the time and publication coverage has helped mitigate the publication bias of outcomes \cite{ampatzoglou2019identifying}.

\subsection{External validity}

Generalizability, referring to the likelihood of not being able to generalize the results, presents an important threat to overcome in SLRs. To address this, we conducted broad searches using one of the most well-known digital libraries (Scopus) to increase the identification of the related primary studies with broad time and publication coverage \cite{ampatzoglou2019identifying}. However, we acknowledge that our findings may not necessarily generalize to grey literature and studies outside the review period.

\subsection{Construct validity}

We are unable to guarantee that we have captured all the relevant primary studies in our SLR. The possibility of missing primary studies is an inevitable limitation in an SLR due to limitations in the search string construction and selection of non-comprehensive digital libraries (DL) \cite{ampatzoglou2019identifying, dybaa2008empirical}. However, to minimise the effects of this, we used several strategies which are described below.

We executed several pilot searches through which we systematically improved the search string to retrieve as many relevant papers as possible. An important point to note is that although the term ``\textit{software security patch management}" is widely used in the industry, this is still a new and emerging topic in research. Thus the use of inconsistent or different terminology in research papers, in particular, the term ``\textit{management}", resulted in a large number of irrelevant studies after its inclusion in the search string. Therefore, we have excluded it from the search string. Although this keyword was not included, the structure of the search string (i.e., broad and not time-bounding) was capable of finding patch management papers, but we had to identify these papers through the study inclusion/exclusion phases. In addition, we used snowballing (i.e., forward and backward search on references of the selected studies) to mitigate the threat of missing relevant primary studies from the exclusion of this term.

Regarding the selection of DLs, while using only Scopus to identify studies may present a limitation of this study, this decision has enabled to increase the coverage of the relevant studies since Scopus is considered the most comprehensive search engine among other digital libraries with the largest indexing system \cite{shahin2017continuous, zahedi2016systematic}. We also did a pilot search on ACM Digital Library to compare and confirm the coverage of results from Scopus. To further mitigate this threat, we made our search string very broad by including the most common keywords to capture as many potentially relevant studies as possible.

\subsection{Conclusion validity}

Researcher bias or the potential bias of authors while interpreting or synthesising the data can impact the conclusions reached \cite{ampatzoglou2019identifying}. To reduce this impact, we adopted the recommended best practices for qualitative data analysis and research synthesis \cite{ampatzoglou2019identifying}. The first author led the data analysis and synthesis and the codebooks were shared with all authors every week where the second and third authors went through all the emergent codes, themes and synthesis results in detail. Disagreements between authors were discussed in detail in weekly meetings until an agreement was reached between all authors.

\section{Conclusion}
\label{section:conclusion}

This study presents our research effort aimed at systematically reviewing and rigorously analysing the literature on software security patch management. We have provided an organised evidential body of knowledge in the area by identifying and categorising the socio-technical challenges and available solutions, and analysing how well the reported solutions have been assessed with their level of rigour and industrial relevance. To the best of our knowledge, this SLR can be considered the first attempt toward systematically reviewing the literature on this topic. Based on a comprehensive analysis of 72 primary studies, we conclude:

\begin{itemize}
    \item The review has enabled us to identify and classify 14 socio-technical challenges and available solutions including 6 themes of \textit{approaches and tools}, and 12 \textit{practices} as common ones affecting all phases of the software security patch management process and those that are specific to each process phase. Moreover, the mapping of challenges onto solutions has revealed that 50\% of the common challenges are not directly associated with solutions unveiling open research challenges for future work.
    
    \item  The distribution of software security patch management solutions is congregated around \textit{vulnerability scanning, assessment and prioritisation}, with 37.5\% of the reviewed studies focused on addressing the challenges related to this phase of the patch management process. In contrast, \textit{patch information retrieval}, \textit{patch testing} and \textit{post-deployment patch verification} have received the least attention with only 5 studies (6.9\%) each implying that there is a need for more studies investigating these underresearched phases.
    
    \item The findings have revealed that only 20.8\% of the solutions have been rigorously evaluated in an industry setting using real-world representative evaluation approaches such as field experiments and case studies. These results indicate a large need for robust evaluation of solutions facilitating industrial relevance, and with more representative and diverse cases.
    
    \item With regard to the research type, a large majority of the reviewed studies (61.1\%) have reported validation research. While only 10 studies (13.9\%) have reported evaluation research, even fewer studies (12.5\%) reported experience papers. The low numbers of evaluation research and experience reports reflect the scarcity of research with industrial relevance. Hence, there is the potential for future studies with active industry collaborations that will result in research outcomes having higher value addition and practical utility.
    
    \item Concerning the reported solution types, 75\% are approaches, 19.4\% are practices and 5.6\% are tools. Further, a large number of limitations in the current tools (e.g., lack of accuracy, security and scalability) have been reported in the reviewed studies. Hence, research and development on new, advanced tools that address the limitations in current tools and support timely software security patch management present a current need. 
    
    \item Even though it has been reported that a significant number of challenges in the software security patch management process emerge from socio-technical aspects such as coordination and collaboration, there is not much empirically known about the role (i.e., how and why) of such socio-technical aspects in the process. Our findings have revealed that the socio-technical aspects have a wide-ranging effect across all phases of the process. Thus we recommend more focus in both research and practice on socio-technical aspects of software security patch management to explore their roles and impact on timely remediation of security vulnerabilities.
    
    \item Despite the widespread attempts to adopt full automation, we note that human-in-the-loop is inevitable in software security patch management due to its inherent complexity and dynamic nature. Based on the findings, we recommend that the emerging research paradigm of \emph{``Human-AI Collaboration"}, which explores how AI-based solutions can be developed to collaborate with human intelligence, presents an important future research opportunity in this topic. 
    
    \item Finally, the mapping of challenges with solutions and the software security patch management process will be beneficial for practitioners to easily understand what approaches, tools, and practices exist for facilitating each challenge. The classification of practices can serve as recommendations for guidance on the successful execution of the software security patch management process. As a direct practical implication of the provided understanding, the security practitioners will be able to identify and assess the factors associated with timely software security patch management and devise suitable decision-making to improve their organisational patching process.

\end{itemize}

\section*{Acknowledgements}

The authors thank the reviewers for their insights and constructive feedback. We also acknowledge the useful feedback provided by our colleagues Bushra Sabir, Chadni Islam and Faheem Ullah on the initial drafts of the paper.
\clearpage
\appendix

\section{Selected primary studies in the review.}
\label{appendix:studieslist}
{\footnotesize\sffamily
\setlength\tabcolsep{2pt}
\centering
\begin{tabularx}{\textwidth}{>{\hsize=0.2\hsize}L
                                  >{\hsize=1.9\hsize}L
                                  >{\hsize=1.2\hsize}L
                                  >{\hsize=1.5\hsize}L
                                  >{\hsize=0.2\hsize}L}
\label{tab:studieslist}\\
    \toprule
\textbf{ID}  & \textbf{Title} & \textbf{Author(s)} & \textbf{Venue} & \textbf{Year} \\
    \midrule
\endfirsthead
   \toprule
\textbf{ID}  & \textbf{Title} & \textbf{Author(s)} & \textbf{Venue} & \textbf{Year} \\
    \midrule
\endhead
    \bottomrule
\endfoot
P1 & Computer security and operating system updates & G. Post, A. Kagan & Information and Software Technology & 2003  \\
\addlinespace
P2 & Reducing internet-based intrusions: Effective security patch management & B. Brykczynski, R.A. Small & IEEE Software & 2003  \\
\addlinespace
P3 & Safe software updates via multi-version execution & P. Hosek, C. Cadar & International Conference on Software Engineering & 2013  \\
\addlinespace
P4 & The attack of the clones: A study of the impact of shared code on vulnerability patching & A. Nappa, R. Johnson, L. Bilge, J. Caballero, T. Dumitraş & IEEE Symposium on Security and Privacy & 2015  \\
P5 & Identifying Information Disclosure in Web Applications with Retroactive Auditing & H. Chen, T. Kim, X. Wang, N. Zeldovich, M. F. Kaashoek & USENIX Symposium on Operating Systems Design and Implementation & 2014  \\
\addlinespace
P6 & Improving VRSS-based vulnerability prioritization using analytic hierarchy process & Q. Liua, Y. Zhanga, Y. Konga, Q. Wu & Journal of Systems and Software & 2012  \\
\addlinespace
P7 & Improving CVSS-based vulnerability prioritization and response with context information & C. Frühwirth and T. Männistö & International Symposium on Empirical Software Engineering and Measurement & 2009  \\
\addlinespace
P8 & Keepers of the Machines: Examining How System Administrators Manage Software Updates & F. Li, L. Rogers, A. Mathur, N. Malkin and M. Chetty & USENIX Conference on Usable Privacy and Security & 2019  \\
\addlinespace
P9 & Towards A Self-Managing Software Patching Process Using Black-Box Persistent-State Manifests & J. Dunagan, R. Roussev, B. Daniels, A. Johnson, C. Verbowski and Y.M. Wang &IEEE International Conference on Autonomic Computing & 2004  \\
\addlinespace
P10 & Security, Availability, and Multiple Information Sources: Exploring Update Behavior of System Administrators & C. Tiefenau, M. Häring, K. Krombholz, E. von Zezschwitz & USENIX Symposium on Usable Privacy and Security & 2020  \\
\addlinespace
P11 & Patch management automation for enterprise cloud & H. Huang, S. Baset, C. Tang, A. Gupta, K.M. Sudhan, F. Feroze, R. Garg, S. Ravichandran & IEEE Network Operations and Management Symposium & 2012  \\
\addlinespace
P12 & Shadow Patching: Minimizing Maintenance Windows in a Virtualized Enterprise Environment & D. Le, J. Xiao, H. Huangy, H. Wang & International Conference on Network and Service Management & 2014  \\
\addlinespace
P13 & VULCAN: Vulnerability Assessment Framework for Cloud Computing & P. Kamongi, S. Kotikela, K. Kavi, M. Gomathisankaran, A. Singhal & International Conference on Software Security and Reliability & 2013  \\
\addlinespace
P14 & VRank: A Context-Aware Approach to Vulnerability Scoring and Ranking in SOA & J. Jiang, L. Ding, E. Zhai, T. Yu & International Conference on Software Security and Reliability & 2012 \\
\addlinespace
P15 & “Anyone Else Seeing this Error?”: Community, System Administrators, and Patch Information & A. Jenkins, P. Kalligeros, K. Vaniea, M. K. Wolters & IEEE European Symposium on Security and Privacy & 2020  \\
\addlinespace
P16 & Staged Deployment in Mirage, an Integrated Software Upgrade Testing and Distribution System & O. Crameri, N. Knezevic, D. Kostic, R. Bianchini, W. Zwaenepoel & ACM SIGOPS Operating Systems Review & 2007  \\
\addlinespace
P17 & Devirtualizable Virtual Machines Enabling General, Single-Node, Online Maintenance & D.E. Lowell, Y. Saito, E.J. Samberg & ACM SIGARCH Computer Architecture News & 2004  \\
\addlinespace
P18 & An automated framework for managing security vulnerabilities & A. Al-Ayed, S.M. Furnell, D. Zhao, P.S. Dowland & Information Management and Computer Security & 2005  \\
\addlinespace
P19 & A cross-site patch management model and architecture design for large scale heterogeneous environment & C.W. Chang, D.R. Tsai, J.M. Tsai & International Carnahan Conference on Security Technology & 2005  \\
\addlinespace
P20 & Security patch management & F.M. Nicastro & Information Systems Security & 2003  \\
\addlinespace
P21 & Intrusion recovery for database-backed web applications & R. Chandra, T. Kim, M. Shah, N. Narula, N. Zeldovich & ACM Symposium on Operating Systems Principles & 2011  \\
\addlinespace
P22 & MAD: A visual analytics solution for Multi-step cyber Attacks Detection & M. Angelini, S. Bonomi, S. Lenti, G. Santucci, S. Taggi & Journal of Computer Languages & 2019  \\
\addlinespace
P23 & Designing an efficient framework for vulnerability assessment and patching (VAP) in virtual environment of cloud computing & R. Patil, C. Modi & Journal of Supercomputing & 2019  \\
\addlinespace
P24 & A new cost-saving and efficient method for patch management using blockchain & Y.Kim, Y. Won & Journal of Supercomputing & 2019  \\
\addlinespace
P25 & Linux patch management: With security assessment features & S. Midtrapanon, G. Wills & International Conference on Internet of Things, Big Data and Security & 2019  \\
\addlinespace
P26 & Vulnus: Visual vulnerability analysis for network security & M. Angelini, G. Blasilli,  T. Catarci, S. Lenti, G. Santucci & IEEE Transactions on Visualization and Computer Graphics & 2018  \\
\addlinespace
P27 & Handling vulnerabilities with mobile agents in order to consider the delay and disruption tolerant characteristic of military networks & T. Aurisch, A. Jacke & International Conference on Military Communications and Information Systems & 2018  \\
\addlinespace
P28 & Green WSUS & S. Kadry, C. Jouma & International Conference on Future Energy, Environment and Materials & 2012  \\
\addlinespace
P29 & Checking running and dormant virtual machines for the necessity of security updates in cloud environments & R. Schwarzkopf, M. Schmidt, C. Strack, B. Freisleben & IEEE International Conference on Cloud Computing Technology and Science & 2011  \\
\addlinespace
P30 & A race for security: Identifying vulnerabilities on 50 000 hosts faster than attackers & M. Procházka, D. Koǔril, R. Wartel, C. Kanellopoulos, C. Triantafyllidis & The International Symposium on Grids and Clouds and the Open Grid Forum & 2011  \\
\addlinespace
P31 & Multi-layered virtual machines for security updates in grid environments & R. Schwarzkopf, M. Schmidt, N. Fallenbeck, B. Freisleben & Euromicro Conference on Software Engineering and Advanced Applications & 2009  \\
\addlinespace
P32 & A study and implementation of vulnerability assessment and misconfiguration detection & C.H. Lin, C.H. Chen, C.S. Laih & IEEE Asia-Pacific Services Computing Conference & 2019  \\
\addlinespace
P33 & Using the vulnerability information of computer systems to improve the network security & Y.P. Lai, P.L. Hsia & Computer Communications & 2007  \\
\addlinespace
P34 & Analyzing enterprise network vulnerabilities & M. Nyanchama, M. Stefaniu & Information Systems Security & 2003  \\
\addlinespace
P35 & The dilemma of security patches & G. Post, A. Kagan & Information Systems Security & 2002 \\
\addlinespace
P36 & Always up-to-date: Scalable offline patching of VM images in a compute cloud & W. Zhou, P. Ning, X. Zhang, G. Ammons, R. Wang, V. Bala & Annual Computer Security Applications Conference & 2010  \\
\addlinespace
P37 & A process framework for stakeholder-specific visualization of security metrics & T. Hanauer, W. Hommel, S. Metzger, D. Pöhn & International Conference on Availability, Reliability and Security & 2018  \\
\addlinespace
P38 & VULCON: A system for vulnerability prioritization, mitigation, and management & K.A. Farris, A. Shah, G. Cybenko, R. Ganesan, S. Jajodia & ACM Transactions on Privacy and Securitys & 2018  \\
\addlinespace
P39 & Patch auditing in infrastructure as a service clouds & L. Litty, D. Lie & International Conference on Virtual Execution Environments & 2011  \\
\addlinespace
P40 & Designing a distributed patch management security system & Y. Nunez,  F. Gustavson, F. Grossman, C. Tappert & International Conference on Information Society & 2010  \\
\addlinespace
P41 & Beyond heuristics: Learning to classify vulnerabilities and predict exploits & M. Bozorgi, L.K. Saul, S. Savage, G.M. Voelker & ACM SIGKDD International Conference on Knowledge Discovery and Data Mining & 2010  \\
\addlinespace
P42 & NetGlean: A methodology for distributed network security scanning & G.W. Manes, D. Schulte, S. Guenther, S. Shenoi & Journal of Network and Systems Management & 2005  \\
\addlinespace
P43 & RL-BAGS: A tool for smart grid risk assessment & Y. Wadhawan, C. Neuman & International Conference on Smart Grid and Clean Energy Technologies & 2018  \\
\addlinespace
P44 & From patching delays to infection symptoms: Using risk profiles for an early discovery of vulnerabilities exploited in the wild & C. Xiao, A. Sarabi, Y. Liu, B. Li, M. Liu, T. Dumitraş & USENIX Security Symposium & 2018  \\
\addlinespace
P45 & PKG-VUL: Security vulnerability evaluation and patch framework for package-based systems & J.H. Lee, S.G. Sohn, B.H. Chang, T.M. Chung & ETRI Journal & 2009  \\
\addlinespace
P46 & A Study of Integrity on the Security Patches System Using PM-FTS & K.J. Kim, M. Kim & Wireless Personal Communications & 2017  \\
\addlinespace
P47 & Patch integrity verification method using dual electronic signatures & J. Kim, Y. Won & Journal of Information Processing Systems & 2017  \\
\addlinespace
P48 & Software asset analyzer: A system for detecting configuration anomalies & X. Li, P. Avellino, J. Janies, M.P. Collins & IEEE Military Communications Conference & 2016  \\
\addlinespace
P49 & Vulnerabilities scoring approach for cloud saas & Z. Li, C. Tang, J. Hu, Z. Chen & International Conference on Ubiquitous Intelligence and Computing, International Conference on Advanced and Trusted Computing, International Conference on Scalable Computing and Communications and its associated Workshops & 2015  \\
\addlinespace
P50 & Risk assessment and mitigation at the information technology companies & B. Marx, D. Oosthuizen & Risk Governance \& Control: Financial Markets and Institutions & 2016  \\
\addlinespace
P51 & A proposed framework for proactive vulnerability assessments in cloud deployments & K.A. Torkura, F. Cheng, C. Meinel & International Conference for Internet Technology and Secured Transactions & 2015  \\
\addlinespace
P52 & Elementary Risks: Bridging Operational and Strategic Security Realms & W. Kanoun, S. Papillon, S. Dubus & International Conference on Signal-Image Technology and Internet-Based Systems & 2015  \\
\addlinespace
P53 & Mining social networks for software vulnerabilities monitoring & S. Trabelsi, H. Plate, A. Abida, M.M. Ben Aoun, A. Zouaoui, C. Missaoui, S. Gharbi, A. Ayari & International Conference on New Technologies, Mobility and Security & 2015  \\
\addlinespace
P54 & A VMM-level approach to shortening downtime of operating systems reboots in software updates & H. Yamada, K. Kono & IEICE Transactions on Information and Systems & 2014  \\
\addlinespace
P55 & A vulnerability life cycle-based security modeling and evaluation approach & G.V. Marconato, M. Kaâniche, V. Nicomette & The Computer Journal & 2013  \\
\addlinespace
P56 & iDispatcher: A unified platform for secure planet-scale information dissemination& M.S. Rahman, G. Yan, H.V. Madhyastha, M. Faloutsos, S. Eidenbenz, M. Fisk & Peer-to-Peer Networking and Applications & 2013  \\
\addlinespace
P57 & Efficient patch-based auditing for web application vulnerabilities & T. Kim, R. Chandra, N. Zeldovich & USENIX Symposium on Operating Systems Design and Implementation & 2012  \\
\addlinespace
P58 & Instant OS updates via userspace checkpoint-and-restart & S. Kashyap, C. Min, B. Lee, T. Kim, P. Emelyanov & USENIX Annual Technical Conference & 2016  \\
\addlinespace
P59 & Tachyon: Tandem execution for efficient live patch testing & M. Maurer, D. Brumley & USENIX Security Symposium & 2012  \\
\addlinespace
P60 & Efficient online validation with delta execution & J. Tucek, W. Xiong, Y. Zhou & International Conference on Architectural Support for Programming Languages and Operating Systems & 2009  \\
\addlinespace
P61 & Enterprise Vulnerability Management and Its Role in Information Security Management & M. Nyanchama & Information Security Management & 2005  \\
\addlinespace
P62 & A Machine Learning-based Approach for Automated Vulnerability Remediation Analysis & F. Zhang, P. Huff, K. McClanahan, Q. Li & IEEE Conference on Communications and Network Security & 2020  \\
\addlinespace
P63 & Reducing Downtime Due to System Maintenance and Upgrades & S. Potter and J. Nieh & USENIX Systems Administration Conference & 2005  \\
\addlinespace
P64 & Increasing virtual machine security in cloud environments & R. Schwarzkopf, M. Schmidt, C. Strack, S. Martin and B. Freisleben & Journal of Cloud Computing: Advances, Systems and Applications & 2012  \\
\addlinespace
P65 & Understanding Software Patching & J. Dadzie & ACM Queue & 2005  \\
\addlinespace
P66 & Patching the Enterprise & G. Brandman & ACM Queue & 2005  \\
\addlinespace
P67 & Why Do Upgrades Fail and What Can We Do about It? & T. Dumitras, P. Narasimhan & ACM/IFIP/USENIX International Conference on Distributed Systems Platforms and Open Distributed Processing & 2009  \\
\addlinespace
P68 & Transparent Mutable Replay for Multicore Debugging and Patch Validation & N. Viennot, S. Nair, J. Nieh & ACM SIGARCH computer architecture news & 2013  \\
\addlinespace
P69 & A quantitative evaluation of vulnerability scanning & H. Holm, T. Sommestad & Information Management \& Computer Security & 2011  \\
\addlinespace
P70 & Evaluation of Security Vulnerability Scanners for Small and Medium Enterprises Business Networks
Resilience towards Risk Assessment & I. Chalvatzis, D. A. Karras, R. C. Papademetriou & IEEE International Conference on Artificial Intelligence and Computer Applications & 2019  \\
\addlinespace
P71 & SLA-driven Applicability Analysis for Patch Management & B. Yang, N. Ayachitula (Arun), S. Zeng, R. Puri & IFIP/IEEE International Symposium on Integrated Network Management & 2011  \\
\addlinespace
P72 & A Design for a Hyperledger Fabric Blockchain-Based Patch-Management System & K. T. Song, S. I. Kim, S. H. Kim & Journal of Information Processing Systems & 2020
\end{tabularx}}

\section{Data Extraction Form.}
\label{appendix:dataextractionform}

{\footnotesize\sffamily
\setlength\tabcolsep{2pt}
\centering
\begin{tabularx}{\textwidth}{>{\hsize=0.2\hsize}L
                                  >{\hsize=1.2\hsize}L
                                  >{\hsize=2.0\hsize}L
                                  >{\hsize=0.6\hsize}L}
\label{tab:dataextractionform} \\
    \toprule
\textbf{ID}  & \textbf{Data item} & \textbf{Description} & \textbf{RQ (Section 3.1)} \\
    \midrule
\endfirsthead
   \toprule
\textbf{ID}  & \textbf{Data item} & \textbf{Description} & \textbf{RQ (Section 3.1)} \\
    \midrule
\endhead
    \bottomrule
\endfoot
D1 & Title & The title of the paper & Demographic data  \\
\addlinespace
D2 & Author(s) & The author(s) of the paper & Demographic data  \\
\addlinespace
D3 & Venue & The publication venue & Demographic data  \\
\addlinespace
D4 & Year & The year of the publication & Demographic data  \\
\addlinespace
D5 & Publication type & The type of publication (e.g., conference paper, journal paper) & Demographic data  \\
\addlinespace
D6 & Area of Focus & The focus of the paper in the software security patch management process & Demographic data  \\
\addlinespace
D7 & Target User(s) & The intended users (e.g., security manager) & Demographic data  \\
\addlinespace
D8 & Application Domain & The target application domain (e.g., cloud) & Demographic data  \\
\addlinespace
D9 & Solution Type & The type of solution i.e., Practice, Approach, Tool & Demographic data  \\
\addlinespace
D10 & Research Type & The type of research i.e., Validation Research, Evaluation Research, Solution Proposal, Experience Paper, Philosophical Paper, Opinion Paper & Demographic data  \\
\addlinespace
D11 & Challenges & The reported socio-technical challenges & RQ1 \\
\addlinespace
D12 & Solutions: Approaches and tools & The proposed approaches and tools (key elements), and their strengths and capabilities documenting how the solution addresses the reported challenges & RQ2.1 \\
\addlinespace
D13 & Solutions: Practices & The reported practices to successfully implement security patch management  & RQ2.2 \\
\addlinespace
D14 & Evaluation & The type of evaluation used to assess the reported solutions, and the level of rigour and industrial relevance & RQ3 \\
\addlinespace
D15 & Limitations and Threats to Validity & The limitations of the solution and the reported threats to validity & Discussion \\
\addlinespace
D16 & Future Work & The reported future work & Discussion \\
\addlinespace
\end{tabularx}}

\bibliography{mybibfile}
\addcontentsline{toc}{section}{References}

\end{document}